\documentclass[%
reprint,
superscriptaddress,
 amsmath,amssymb,amsthm,
 aps,
 prx
floatfix,
]{revtex4-2}

\usepackage{graphicx}
\usepackage{dcolumn,amsthm}
\usepackage{bm}
\usepackage{ulem}

\usepackage{txfonts}
\usepackage{dsfont}
\usepackage[english]{babel}

\usepackage[colorlinks=true,linkcolor=blue,citecolor=blue,urlcolor=blue]{hyperref}

\newcommand{\ie}{\textit{i.e.,\ }}

\newcommand{\<}{\langle}
\renewcommand{\>}{\rangle}
\newcommand{\abs}[1]{\left\lvert#1\right\rvert}
\newcommand{\norm}[1]{\lVert#1\rVert}
\newcommand{\A}{\mathcal{A}}

\newcommand{\1}{\mathds{1}}

\providecommand{\tr}{{\rm Tr}}
\renewcommand{\phi}{\varphi}
\newcommand{\bra}[1]{\langle #1\rvert}
\newcommand{\ket}[1]{\lvert #1\rangle}

\renewcommand{\Re}{\mathbb{R}}

\begin{document}

\title{Homodyne detection of non-Gaussian quantum steering}

\author{Carlos E. Lopetegui}
\affiliation{Laboratoire Kastler Brossel, Sorbonne Universit\'{e}, CNRS, ENS-Universit\'{e} PSL,  Coll\`{e}ge de France, 4 place Jussieu, F-75252 Paris, France}
\affiliation{Laboratoire de Physique de l’Ecole Normale Sup\'erieure, ENS, Universit\'e PSL, CNRS, Sorbonne Universit\'e, Universit\'e de Paris, F-75005 Paris, France}
\author{Manuel Gessner}
\affiliation{ICFO-Institut de Ci\`{e}ncies Fot\`{o}niques, The Barcelona Institute of Science and Technology, Av. Carl Friedrich Gauss 3, 08860, Castelldefels (Barcelona), Spain}
\affiliation{Laboratoire Kastler Brossel, Sorbonne Universit\'{e}, CNRS, ENS-Universit\'{e} PSL,  Coll\`{e}ge de France, 4 place Jussieu, F-75252 Paris, France}
\author{Matteo Fadel}
\affiliation{Department of Physics, ETH Z\"{u}rich, 8093 Z\"{u}rich, Switzerland}
\author{Nicolas Treps}
\affiliation{Laboratoire Kastler Brossel, Sorbonne Universit\'{e}, CNRS, ENS-Universit\'{e} PSL,  Coll\`{e}ge de France, 4 place Jussieu, F-75252 Paris, France}
\author{Mattia Walschaers}
\email{mattia.walschaers@lkb.upmc.fr}
\affiliation{Laboratoire Kastler Brossel, Sorbonne Universit\'{e}, CNRS, ENS-Universit\'{e} PSL,  Coll\`{e}ge de France, 4 place Jussieu, F-75252 Paris, France}

\date{\today}

\begin{abstract}
Quantum correlations are at the core of current developments in quantum technologies. Certification protocols of entanglement and steering, suitable for continuous-variable non-Gaussian states are scarce and generally highly demanding from an experimental point of view. We propose a protocol based on Fisher information for witnessing steering in general continuous-variable bipartite states, through homodyne detection. It proves to be relevant for the detection of non-Gaussian steering in scenarios where witnesses based on Gaussian features like the covariance matrix are shown to fail.  
\end{abstract}

\maketitle


\section{\label{sec:Intro} Introduction}
In 1935 Einstein, Podolsky and Rosen introduced what came to be known as the EPR paradox \cite{EPR1935}, challenging, through the argument of local realism the completeness of quantum mechanics. In his early response \cite{Schrod1935,Schrod1936}, Schr\"odinger addressed the issue of spooky action, troubled by the paradox arising from the capability of one part of a bipartite system to instantaneously \textit{steer} the state of the other through appropriate local measurements. These works received notorious attention after the seminal paper by Bell \cite{Bell1964}, who proposed a strong test for locality itself. In 2007, Wisemann et al. \cite{Wiseman2007} provided an operational benchmark for steering, from which they proved that the set of states that manifest steering are a strict subset of the set of entangled states and a strict super-set of those which violate Bell inequalities. This definition can be understood in terms of a scenario where two parties, Alice and Bob, share a state. Alice has to convince Bob that the state they share is entangled, while Bob does not actually trust Alice, \ie he does not assume her measurements to be in accordance with the constraints imposed by quantum physics. Alice will communicate the results of her measurements and then Bob can measure the state on his part of the system. Whenever Bob can verify the presence of a quantum correlation based only on the information provided by Alice and his own measurement results, we say that there was quantum steering from Alice to Bob. 

The relevance of the characterization of steering goes beyond the interest in fundamental questions as it 
is a relevant resource in quantum information protocols \cite{Uola2020,Aolita2015}, like one-sided device independent quantum key distribution \cite{Branciard2012,Gehring2015, Walk2016}, certification of random number generators \cite{Law_2014,Passaro_2015}, quantum metrology \cite{EPR_Metr2021}, and quantum channel discrimination \cite{Piani2015}. These one-sided device independent approaches to quantum information protocols are settled in between the fully device independent protocols, that require the violation of Bell inequalities for certification, and the entanglement-based protocols, which are less restrictive, but also slightly less secure \cite{PhysRevLett.98.230501,RevModPhys.86.419,Horodecki2009}. 

The problem of steering characterization for Gaussian states has been widely studied \cite{Reid2009,Reid2009Colloquium}, and a well-defined measure has been established \cite{Adesso2015,PhysRevLett.117.220502,8004445}, based on the symplectic spectrum of the conditioned covariance matrix. 
However, for many applications in quantum technologies, one requires non-Gaussian states. For example, non-Gaussian features are necessary to reach a quantum computational advantage \cite{Mari2012}, and for quantum error correction \cite{PhysRevLett.102.120501}. Any application that relies on entanglement distillation must be non-Gaussian \cite{PhysRevLett.89.137903} and common entanglement distillation protocols effectively create non-Gaussian quantum correlations \cite{EntDistil,PhysRevLett.98.030502}. Such non-Gaussian quantum correlations become particularly relevant in quantum metrology, where they often lead to an improvement in sensitivity \cite{PhysRevLett.102.100401,PhysRevLett.122.090503,QMetReview,RevModPhys.90.035005}.

A general characterization of steering in non-Gaussian scenarios, has been elusive so far. One possible approach relies on conditional quantum state tomography and semidefinite programming \cite{PhysRevLett.121.170403}. Alternatively, many protocols are based on second order correlations \cite{Reid2009}, and for non-Gaussian states these protocols require non-Gaussian measurements \cite{PhysRevA.96.042326}. The latter is twofold undesired: First, it is appealing to rely strictly on Gaussian continuous-variable (CV) measurements, such as homodyne detection. Second, we want to probe the non-Gaussian features of the state, and thus must avoid introducing any additional non-Gaussian features through the measurement. In this spirit, we aim for a general protocol purely based on homodyne detection. Even though methods based on hierarchies have been proposed \cite{PhysRevLett.115.210401}, these can require significant experimental and computational overhead when high-order moments are involved. Thus, rather than only focusing on moments of the measurement outcomes, our protocol will exploit the full measurement statistics. 

We tackle the problem of witnessing quantum steering with a toolbox based on quantum metrology \cite{Giovannetti2011,Toth2014,Pezze2018}. The steering capacity in a bipartite system was formally linked to an enhancement in the capability to estimate certain parameters \cite{EPR_Metr2021}.
We adapt this approach to the experimental context and limitations of CV quantum optics and show its relevance for non-Gaussian states. For that, we will consider single-photon-subtracted states as a probe system. In the context of non-Gaussian states, photon subtraction, offers an experimentally feasible way to attain Wigner negativity in a controlled way \cite{Tutorial2021,Chabaud2021}. This approach offers a very flexible way to generate different kind of states \cite{Ra2020} and in particular purely non-Gaussian features can be studied by appropriately choosing the mode in which the photon is subtracted \cite{Tutorial2021}. These states are relevant probe since pure photon-subtracted squeezed vacuum states have been shown to manifest quantum steering that cannot be detected by variance-based criteria \cite{PhysRevA.89.012104}. We also show that our metrological approach detects more non-Gaussian steerable states than the entropic criterion of \cite{PhysRevLett.106.130402}, even though the latter also exploits full homodyne statistics.

\section{\label{sec:Fisher} Protocol}
\subsection{Protocol for general quantum states}\label{sec:YFG}

We will now formulate the steering detection scheme as a metrological protocol, following \cite{EPR_Metr2021}. We consider the scenario in which Bob attempts to estimate a phase $\xi$ generated by a Hamiltonian $\hat H$ that acts on his side of the system. Without any further information than the one he can extract from direct measurements in the displaced state $\hat \rho^{B}_{\xi}=\exp(-i\xi \hat H)\hat \rho^{B} \exp(i\xi \hat H)$, the maximal precision that he can achieve using an arbitrary unbiased estimator $\xi_{est}$ is limited by the quantum Fisher Information (QFI) $F_Q(\hat \rho^{B},\hat H)$, the central quantity in quantum metrology \cite{Giovannetti2011,Toth2014,Pezze2018}. In the present scenario, where the parameter to be estimated is implemented by a unitary transformation, generated by a Hamiltonian, there is a practical expression for the QFI for a state $\hat \rho^B = \sum_k r_k \ket{r_k}\bra{r_k}$:
\begin{equation}
    F_Q(\hat \rho^{B},\hat H) = 4 \tr [\hat \rho^B \hat H^2] - 8 \sum_{j,k} \frac{r_k r_j}{r_k + r_j} \abs{\bra{r_j}\hat H\ket{r_k}}^2.
\end{equation}
Note that this expression requires us to know the eigenvalues $r_k$ and associated eigenvectors $\ket{r_k}$. However, in many physical systems, and notably CV systems where the density matrix is infinite dimensional, these quantities are often not known. 

The QFI represents the sensitivity of the state $\hat{\rho}^B$ under small perturbations generated by $\hat{H}$. This idea is formalised in the quantum Cram\'er-Rao bound on the variance of the estimator
\begin{equation}\label{eq:CRB}
\text{Var}(\xi_{est})\geq \frac{1}{n F_Q(\hat \rho^B,\hat H)},
\end{equation}
where $n$ is the number of repetitions of the measurement protocol. The inequality can be saturated by choosing the optimal measurement observable and estimator. 

Nevertheless, Bob's state might be correlated with another system. Let us assume that Alice possesses this second party, and will assist Bob in his estimation protocol by sending him information about her measurement setup and outcome. Alice's assistance may improve Bob's estimation precision even when correlations are purely classical. Local complementarity sets a limit to this improvement that can only be overcome when there is quantum steering \cite{EPR_Metr2021}.  
The average sensitivity attainable by Bob following assistance by Alice, is upper-bounded by the conditional QFI
\begin{equation}\label{eq:Cond_QFI}
F_{Q}^{B|A}(\mathcal{A},\hat H)\coloneqq\max_{\hat X}\int p(a|\hat X) F_{Q}^{B}(\hat \rho^{B}_{a\lvert\hat X},\hat H)da
\end{equation}
and we introduce the assemblage as a function $\mathcal{A}$ that maps the observable $\hat X$ and one of its measurement outcomes $a$ to  
\begin{equation}\label{eq:Assemb}
\mathcal{A}(a,\hat X) \coloneqq p(a|\hat X) \hat \rho^{B}_{a\lvert\hat X} , 
\end{equation}
where $p(a|\hat X)$ is the probability distribution for Alice's outcomes $a$ after measurement of the observable $\hat X$, and $\hat \rho^{B}_{a\lvert\hat X}$ the conditioned state on Bob's side that is obtained after such a measurement. 

In this context the confirmation of quantum steering consists in showing that the assemblage \eqref{eq:Assemb} cannot be described with a hidden state model given by
\begin{equation}\label{eq:LHS}
\mathcal{A}(a,\hat X)=\int d\lambda p(\lambda) p(a|\hat X,\lambda) \hat \sigma_{\lambda}^{B}.
\end{equation}
Note, moreover, that the implementation of a local phase $\xi$ preserves the structure of the local hidden state model. If the state Bob and Alice share is consistent with the structure of \eqref{eq:LHS}, the following inequality holds \cite{EPR_Metr2021}
\begin{equation}\label{eq:Metrological_Inequality}
F_{Q}^{B|A}(\mathcal{A},\hat H)\leq 4 \text{Var}_{Q}^{B|A}(\mathcal{A},\hat H),    
\end{equation}
where $\text{Var}_{Q}^{B|A}(\mathcal{A},\hat H)$ represents the quantum conditional variance 
\begin{equation}\label{eq:condVar}
\text{Var}_{Q}^{B|A}(\mathcal{A},\hat H) \coloneqq \min_{\hat X}\int p(a|\hat X) \text{Var}(\hat \rho_{a\lvert\hat X}^{B},\hat H)da,
\end{equation}
that is obtained after minimization over all possible measurement setups by Alice. Here we encounter the variance of $\hat H$ in the state $\rho_{a\lvert\hat X}^{B}$, given by
\begin{equation}
\text{Var}(\hat \rho_{a\lvert\hat X}^{B},\hat H) \coloneqq {\rm Tr}[\hat \rho_{a\lvert\hat X}^{B} \hat H^2] - {\rm Tr}[\hat \rho_{a\lvert\hat X}^{B} \hat H]^2.
\end{equation}
Together with the Cram\'er-Rao bound, \eqref{eq:Metrological_Inequality} implies the uncertainty relation \cite{EPR_Metr2021}
\begin{equation}
    \text{Var}(\xi_{est}) \text{Var}_{Q}^{B|A}(\mathcal{A},\hat H) \geq \frac{1}{4n},
\end{equation}
between the phase displacement estimator $\xi_{est}$ and its generator $\hat H$, whose violation constitutes an EPR paradox.

Inequality~(\ref{eq:Metrological_Inequality}) can be thought of as a way to witness steering through its relevance for metrological tasks. The extent to which a given assemblage violates the inequality is captured by the steering witness
\begin{equation}\label{eq:Steering_witness}
    S_{\text{max}}(\mathcal{A})=\max_{\{\hat H,\mathrm{Tr}(\hat H^2)=1\}}\left[F_{Q}^{B|A}(\mathcal{A},\hat H)-4 \text{Var}_{Q}^{B|A}(\mathcal{A},\hat H)\right]^{+},
\end{equation}
where $\left[x\right]^{+}=\max\{0,x\}$. Moreover, Reid's criterion \cite{PhysRevA.40.913} can be derived as a weaker version of this witness. It can be shown \cite{EPR_Metr2021} that 
\begin{equation}\label{eq:FI_lower_bound}
F^{B|A}_{Q}(\mathcal{A}, \hat H)\geq \frac{\abs{\<\left[ \hat H,\hat{M}\right]\>_{\hat \rho^{B}}}^2}{\text{Var}_{Q}^{B|A}\left(\A,\hat{M}\right)}
\end{equation}
holds for arbitrary assemblages $\mathcal{A}$ and observables $\hat H$ and $\hat{M}$. Combined with \eqref{eq:Metrological_Inequality}, we introduce the following measure for the violation of Reid's variance-based steering witness
\begin{equation}\label{eq:Reid}
S_R({\cal A}) = \max_{\{\hat H,\mathrm{Tr}(\hat H^2)=1\}} \left[ \frac{\abs{\<\left[ \hat H,\hat{M}\right]\>_{\hat \rho^{B}}}^2}{\text{Var}_{Q}^{B|A}\left(\mathcal{A},\hat{M}\right)} - 4 \text{Var}_{Q}^{B|A}(\mathcal{A},\hat H)\right]^+.\end{equation}
This witness is very commonly used to witness steering in Gaussian states with quadrature operators \cite{PhysRevA.40.913}. Furthermore, \eqref{eq:FI_lower_bound} directly implies that $S_{\text{max}}({\cal A}) \geqslant S_R({\cal A})$.

\subsection{Homodyne protocol for continuous-variable systems}\label{sec:ProtocolCV}
In this Section, we translate the general protocol of the previous Section to the specific context of multimode quantum optics \cite{Treps2020,Tutorial2021}.  We rely on quadrature displacements as the phase estimation probe, which can be easily implemented by shifting the Wigner function \cite{Wigner1932} in phase space. Experimentally, such a displacement results in a simple shift of the measured quadrature histograms, which implies that the effect of the parameter can easily be ``simulated'' in post-processing. This will allow us to develop a framework to witness steering based entirely on homodyne detection.\\

Our starting point is the $M$-mode electric field operator
\begin{equation}
\hat{E}^{+}(\mathbf{r},t)=\sum^M_{j=1} \epsilon_{j} \hat{a}_j u_j(\mathbf{r},t)    
\end{equation}
where $u_i(\mathbf{r},t)$ are a set of orthonormal solutions of Maxwell equations (classical modes), $\epsilon_j$ is a constant that carries the dimensions of the field, and $\hat{a}_j$ are the annihilation operators corresponding to modes $u_j$ of the bosonic field. 
In CV quantum optics the fundamental observables are the real  and complex components of these operators, defined as
\begin{equation}
\hat{a}_j=\frac{\hat{q}_{j}+i \hat{p}_j}{2},
\end{equation}
where $\hat{q}_j$ and $\hat{p}_j$ are the amplitude and phase quadratures of the electric field, respectively, which satisfy the canonical commutation relation
$\left[\hat{q}_j,\hat{p}_k\right]=2 i \delta_{j,k}$.  The measurement outcomes for these observables are represented in the optical phase space, which has a symplectic structure associated to the form
\begin{equation}
    \Omega = \bigoplus_{j=1}^M \begin{pmatrix}0 & -1 \\
    1 & 0\end{pmatrix}.
\end{equation}
We can now define vectors of quadrature operators \begin{equation}\vec{\hat x} = (\hat q_1, \hat p_1, \dots, \hat q_M, \hat p_M)^{\top},\end{equation} 
and translate the commutation relation to $[\hat x_j, \hat x_k] = 2i \Omega_{jk}$.

To represent quantum states on optical phase space we resort to a quasi-probability distribution, the Wigner function. Even though this representation can reach negative values and is thus not a joint probability distribution for quadratures, its marginals describe the probabilities of measurement outcomes for individual quadrature observables \cite{Tutorial2021}. We will focus on states of a bipartite system that are completely described by its Wigner function $W(\vec{x}_{A}\oplus \vec{x}_{B})$ in a phase space of dimension $\Re^{2m}\oplus \Re^{2m'}$, where $\vec{x}_{A(B)}$, stand for the phase-space coordinates of subsystem $A(B)$ which consists of $m(m')$ modes.\\ 

A direct application of the protocol in Section \ref{sec:YFG} would require us to obtain the QFI $F_{Q}^{B}$. This is in general a notoriously difficult task as it involves the reconstruction of the density matrix, which is often unfeasible in a CV setting. 
However, the QFI is lower bounded by its classical counterpart
\begin{equation}\label{eq:QFI-FI}
F_{Q}^{B}(\hat \rho^{B},\hat H)\geq F^B_{\xi}[P].
\end{equation}
The classical FI characterizes the best precision that can be obtained for estimating $\xi$ by using the results of a specific measurement. It is defined as
\begin{equation}\label{eq:Classical_FI}
    F^B_{\xi}[P]\coloneqq \int_{\Re} P(q|\xi) \left(\frac{\partial \mathcal{L}(q|\xi)}{\partial \xi}\right)^2 d q
\end{equation}
where $\mathcal{L}(q|\xi)=\log[P(q|\xi)]$ represents the logarithmic likelihood associated to the probability density of measurement outcomes $q$, after implementation of the parameter $\xi$. More formally phrased, $P(q|\xi) = {\rm Tr}[\hat \rho^{B}_{\xi} \hat \Pi_q]$, where $\hat \Pi_q$ form a positive operator-valued measure (POVM) such that $ \int \hat\Pi_q dq = \mathds{1}$. For CV systems, it is natural to choose $\hat H$ to be a quadrature operator, and $\hat\Pi_q = \ket{q}\bra{q}$ to correspond to homodyne measurements.

The relation \eqref{eq:QFI-FI} is particularly appealing as it shows that any violation of the inequality \eqref{eq:Metrological_Inequality} based on the classical FI, is a lower bound for the exact violation based on the QFI. The down-side of relying on the classical FI is that one may fail to witness steering that could otherwise be detected by using a better measurement scheme. However, the classical Fisher information already provides a strict improvement over Reid's criterion \eqref{eq:Reid}. We will show that this improvement is sufficient to witness non-Gaussian steering.\\

\begin{figure}[!htb]
\centering
\includegraphics[width=0.48\textwidth]{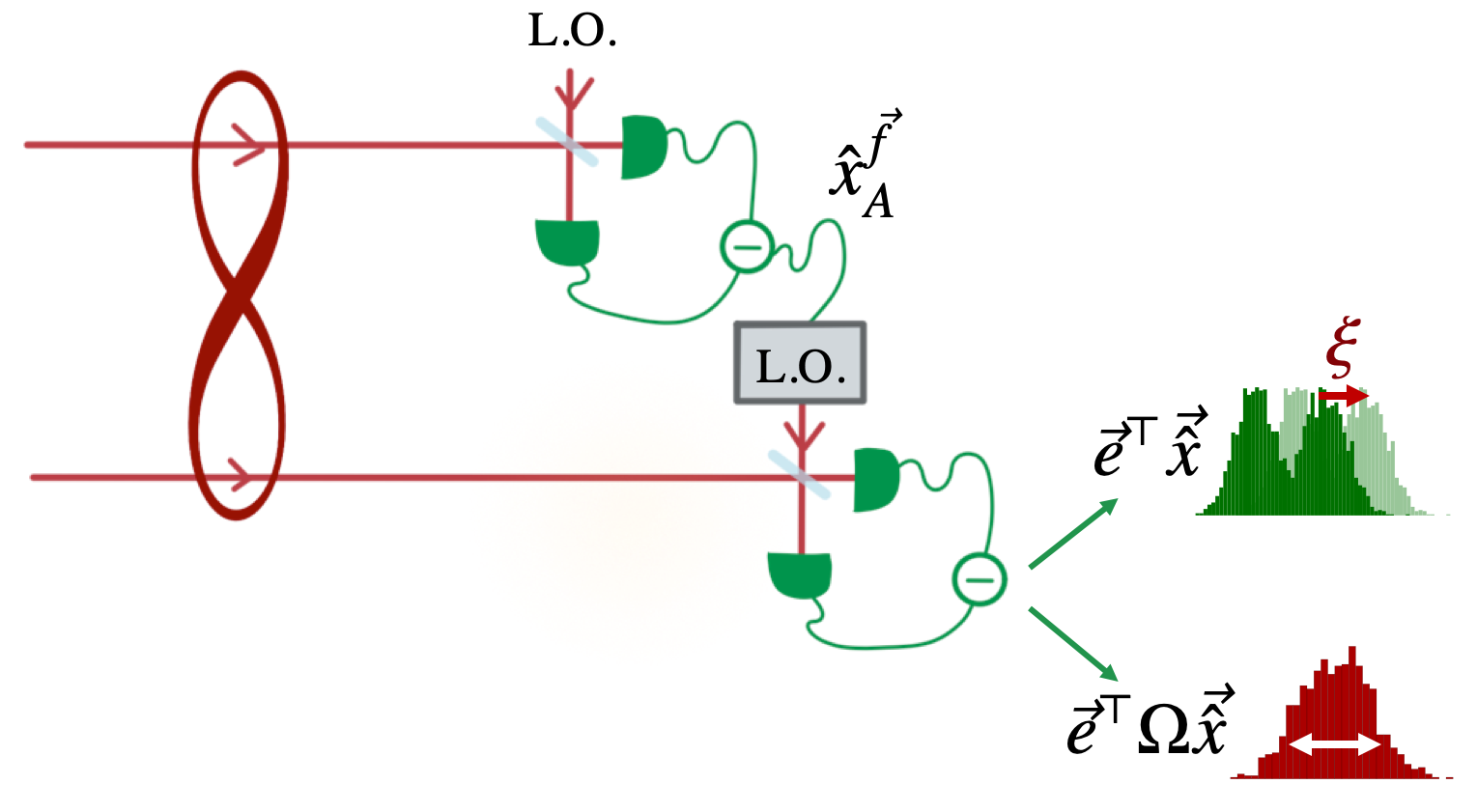}
\caption{Metrological protocol on which we base the witnessing of steering for bipartite CV states. Alice performs homodyne detection on the mode she owns and communicates to Bob the quadrature she chose to measure and its outcome. Based on this information Bob chooses what quadrature to measure in order to better estimate the displacement $\xi$ generated by $\hat D(\xi) = \exp\left[-i \xi\, {\vec e}^{\,T} \Omega \vec{\hat x}/2\right]$, such that the Hamiltonian is given by $\hat H={\vec e}^{\,T} \Omega \vec{\hat x}/2$.}
\label{fig:Metrological_Protocol}
\end{figure}

In what follows we summarize the protocol to witness steering for a bipartite CV system, see Fig.\ref{fig:Metrological_Protocol}. We have two sets of modes, in principle, mutually entangled, one in possession of Alice and one in possession of Bob. In her modes, Alice performs a homodyne detection that is characterized by a normalized vector $\vec f$ in Alice's phase space, which means she measures the quadrature $\hat x_A^{\vec f} = \vec f^{\top} \vec{\hat x}$. When she obtains the measurement result $x_0$, Bob's state will be transformed into a state described by the conditional Wigner function
\begin{equation}\label{eq:WignerCond}
\begin{split}
    &W^{B|A}(\vec{x}_{B}|x_{A}^{\vec f}=x_0)\\
    &=\frac{\int_{\Re^{2m}} W(\vec{x}_{A}\oplus \vec{x}_{B}) \delta(\vec f^{\top}\vec x_A - x_0) d\vec{x}_{A}}{\int_{\Re^{2m}\oplus \Re^{2m'}}W(\vec{x}_{A}\oplus \vec{x}_{B}) \delta(\vec f^{\top}\vec x_A-x_0) d\vec{x}_{A} d\vec{x}_{B}}.
    \end{split}
\end{equation}
Bob estimates a local quadrature displacement $W^{B|A}(\vec{x}_{B})\mapsto W^{B|A}(\vec{x}_{B}-\xi \vec e)$ on his subsystem. The parameter of interest $\xi$ here corresponds to the extent of this displacement, which is generated by the Hamiltonian $\hat H = \vec e^{\top} \Omega \vec{\hat x}/2$, with $\vec e$ a normalised vector in Bob's phase space. 
In the spirit of \eqref{eq:Steering_witness}, to witness steering we optimize over all possible choices of displacement axis, and thus maximize over $\vec e$. 

To study Bob's sensitivity for such an estimation, we evaluate the quantities involved in the inequality \eqref{eq:Steering_witness}, but we will replace the QFI with the classical FI \eqref{eq:Classical_FI}. To compute the classical FI we fix the observable $\hat M$. A logical choice is to measure the displaced quadrature, given by $\hat M = \vec e^{\top} \vec {\hat x}$. This means that $P(q\mid \xi)$ in \eqref{eq:Classical_FI} is the marginal of the Wigner function \eqref{eq:WignerCond} along the phase space axis $\vec e$. The probability of obtaining an outcome $q$ when measuring the quadrature along $\vec e$ is given by
\begin{equation}\label{eq:wahwah}
    P^{B}_{x_0 \lvert \vec f}(q)=\int_{ \mathbb{R}^{2m'}}  \delta( \vec e^{\top}\vec{x}_B - q)W^{B|A}(\vec{x}_{B}|x_{A}^{\vec f}=x_0) d\vec{x}_{B}.
\end{equation}
The displaced profile is obtained by the map $q\mapsto q-\xi$ on the marginal distribution, such that we can write
\begin{equation}\label{eq:PB}P^{B}_{x_0 \lvert \vec f}(q|\xi)=P^{B}_{x_0 \lvert \vec f}(q-\xi).
\end{equation}
The resulting conditional classical FI for a fixed choice of Bob's displacement and measurement (determined by $\vec{e}$), optimized over all homodyne observables ($\vec{f}$) on Alice's side is defined as
\begin{equation}\label{eq:class_FI_hom}
    F^{B|A}_{\rm hom}\left(\mathcal{A},\frac{\vec e^{\top} \Omega \vec{\hat x}}{2}\right)=\max_{\vec f \in \Re^{2m} }\int_{\Re} P_A(x_A^{\vec f}=x_0) F^{B}_{\xi}[P^{B}_{x_0 \lvert \vec f}]dx_0.
\end{equation}
Here, $P_A(x_A^{\vec f}=x_0)$ is the marginal of the Wigner function along the quadrature measured by Alice. To check whether there is some mode in Alice's subsystem that can steer Bob's the optimization runs over all possible choices of $\vec f$. One could refine the question and restrict $\vec f$ to the phase space of one specific mode to test whether this particular mode can steer Bob's subsystem.

To compute the conditional variance of the generator $\vec e^{\top} \Omega \vec{\hat x}/2$, we also use a marginal of the conditional Wigner function \eqref{eq:WignerCond}. From definition \eqref{eq:condVar}, we find that the conditional variance is given by
\begin{equation}\begin{split}
&\text{Var}_{\rm hom}^{B|A}\left(\mathcal{A},\frac{ \vec e^{\top} \Omega \vec{\hat x}}{2}\right)\\&=\min_{\vec f \in \Re^{2m}}\frac{1}{4}\int_{\Re}P_A(x_A^{\vec f}=x_0) \text{Var}\left(\hat \rho_{x_0 \lvert \vec f}^{B},\vec e^{\top} \Omega \vec{\hat x}\right)dx_0,
\end{split}
\end{equation}
where $\text{Var}(\hat \rho_{x_0 \lvert \vec f}^{B},\vec e^{\top} \Omega \vec{\hat x})$ is the variance of the quadrature corresponding to the generator $\vec e^{\top} \Omega \vec{\hat x}$. To compute this quantity, we introduce the probability of obtaining an outcome $p$ when we measure the quadrature along the axis $\Omega \vec e$

\begin{equation}\label{eq:cond_mom_prob_dens}
    \tilde{P}^{B}_{x_0 \lvert \vec f}(p)=\int_{\mathbb{R}^{2m'}}  \delta(\vec e^{\top}\Omega\vec{x}_{B} - p)W^{B|A}(\vec{x}_{B}|x_{A}^{\vec f}=x_0) d\vec{x}_{B},
\end{equation}
This distribution allows us to compute
\begin{equation}\label{eq:var}
\text{Var}\left(\hat \rho_{x_0 \lvert \vec f}^{B},\vec e^{\top} \Omega \vec{\hat x}\right) =\int_{\Re} p^2 \tilde{P}^{B}_{x_0 \lvert \vec f}(p)\, dp - \left(\int_{\Re} p \tilde{P}^{B}_{x_0 \lvert \vec f}(p)\, dp\right)^2
\end{equation}
In other words, Alice first chooses a mode and quadrature to measure. Bob then also chooses a mode and a quadrature to measure depending on Alice's choice. Alice communicates her measurement outcomes to Bob, and Bob will group his measurement outcomes depending on Alice's result.

Finally, in analogy to the general definition \eqref{eq:Steering_witness}, which optimizes over all Hamiltonians, we optimize our homodyne steering witness over all possible displacement vectors. This leads to the final witness
\begin{equation}\label{eq:Steering_witness_hom}\begin{split}
    S_{\text{max}}^{\text{hom}}(\mathcal{A})=\max_{\vec e \in \Re^{2m'}}\left[F^{B|A}_{\rm hom}\left(\mathcal{A},\frac{\vec e^{\top} \Omega \vec{\hat x}}{2}\right)-\text{Var}_{\rm hom}^{B|A}\left(\mathcal{A},\vec e^{\top} \Omega \vec{\hat x}\right)\right]^{+},
    \end{split}
\end{equation}
for quantum steering with the specialized homodyne-based protocol. Note that we have used that $4\text{Var}_{\rm hom}^{B|A}\left(\mathcal{A},\vec e^{\top} \Omega \vec{\hat x}/2\right) = \text{Var}_{\rm hom}^{B|A}\left(\mathcal{A},\vec e^{\top} \Omega \vec{\hat x}\right)$.

Even though our protocol is formulated in a fully multimode context, it will effectively detect quantum steering between two optical modes, one given by $\vec f$ on Alice's side and one given by $\vec e$ on Bob's system. Optimising over the possible choices of $\vec f$ and $\vec e$ gives us a sufficient criterion for steering from Alice to Bob, but one can make the protocol more general by measuring multiple quadratures simultaneously on, both, Alice's and Bob's side of the system. Because this extension is technically rather involved, but physically straightforward, we present it separately in Appendix \ref{app:MultimodeMode}.

The witness \eqref{eq:Steering_witness_hom} for our homodyne-based protocol is a lower bound for the steering witness proposed in \cite{EPR_Metr2021} that relies on the QFI. At the same time, we can define a version of Reid's criterion \eqref{eq:Reid} restricted to homodyne measurements by setting $\hat H = \vec e^{\top} \Omega \vec {\hat x}$ and $\hat M = \vec e^{\top} \vec {\hat x}$, which leads to
\begin{equation}\label{eq:ReidHom}
    S_{R}^{\rm hom} (\mathcal{A})=\max_{\vec e \in \Re^{2m'}}\left[\frac{1}{\text{Var}_{\rm hom}^{B|A}\left(\A,\vec e^{\top} \vec {\hat x}\right)}-\text{Var}_{\rm hom}^{B|A}\left(\mathcal{A},\vec e^{\top} \Omega \vec{\hat x}\right)\right]^{+}.
\end{equation}
Here, we find the quantity  $\text{Var}_{\rm hom}^{B|A}\left(\A,\vec e^{\top} \vec {\hat x}\right)^{-1}$
which quantifies the sensitivity of estimating $\xi$ based only on the average measurement outcome of $\vec e^{\top} \vec {\hat x}$. Due to the relation between the method of moments and the Fisher information \cite{RevModPhys.90.035005}, this is always smaller than the sensitivity set by the FI. We thus find the hierarchy $S_{R}^{\rm hom}(\A)\leq S_{\max}^{\text{hom}}(\A)\leq S_{\text{max}}(\A)$. Interestingly, there are states for which $S_{\max}^{\text{hom}}(\A) < S_{R}(\A)$ as the general version of Reid's criterion allows for highly non-Gaussian operators $\hat H$ and $\hat M$.

Finally, it is interesting to explicitly compare $S_{R}^{\rm hom} (\mathcal{A})$ and $S^{\rm hom}_{\rm max}(\mathcal{A})$ for Gaussian states. When Alice conditions on a homodyne measurement, she performs a Gaussian operation on the state. When the global state is Gaussian, Alice's measurement will create a Gaussian conditional state $\rho_{x_0 \lvert \vec f}^{B}$ on Bob's subsystem \cite{RevModPhys.84.621}. Because the state is Gaussian, it is characterized by a Gaussian Wigner functions and its marginals are also Gaussian. Therefore, the probability distribution $P^{B}_{x_0 \lvert \vec f}(q|\xi)$ in \eqref{eq:PB} is Gaussian and only its mean value depends on the parameter $\xi$. In this case, a simple calculation shows that $F^{B}_{\xi}[P^{B}_{x_0 \lvert \vec f}] = 1/\text{Var}\left(\hat \rho_{x_0 \lvert \vec f}^{B},\vec e^{\top} \vec{\hat x}\right)$. This leads us to the following identity for Gaussian states 
\begin{equation}
    F^{B|A}_{\rm hom}\left(\mathcal{A},\frac{\vec e^{\top} \Omega \vec{\hat x}}{2}\right)=\max_{\vec f \in \Re^{2m} }\int_{\Re} P_A(x_A^{\vec f}=x_0) \frac{1}{\text{Var}\left(\hat \rho_{x_0 \lvert \vec f}^{B},\vec e^{\top} \vec{\hat x}\right)}dx_0.
\end{equation}
A second important element for Gaussian states is that $\text{Var}\left(\hat \rho_{x_0 \lvert \vec f}^{B},\vec e^{\top} \Omega \vec{\hat x}\right)$ is independent of actual measurement result $x_0$ on Alice's side \cite{RevModPhys.84.621}. In other words, we find that
\begin{equation}
    F^{B|A}_{\rm hom}\left(\mathcal{A},\frac{\vec e^{\top} \Omega \vec{\hat x}}{2}\right)=\max_{\vec f \in \Re^{2m}}\frac{1}{\text{Var}\left(\hat \rho_{x_0 \lvert \vec f}^{B},\vec e^{\top} \vec{\hat x}\right)}.
\end{equation}
From the same argument, it follows that 
\begin{equation}
    \frac{1}{\text{Var}_{\rm hom}^{B|A}\left(\A,\vec e^{\top} \vec {\hat x}\right)} = \max_{\vec f \in \Re^{2m}}\frac{1}{\text{Var}\left(\hat \rho_{x_0 \lvert \vec f}^{B},\vec e^{\top} \vec{\hat x}\right)},
\end{equation}
which ultimately shows that 
\begin{equation}
    S_{R}^{\rm hom} (\mathcal{A}) = S^{\rm hom}_{\rm max}(\mathcal{A}) \quad \text{ for Gaussian states.}
\end{equation}
This shows that our metrological formalism based on quadrature measurements can only outperform Reid's criterion based on quadrature variances when we are dealing with non-Gaussian states.\\

Reid's criterion as captured by $S_{R}^{\rm hom} (\mathcal{A})$ is also a lower bound for a different steering witness that can be derived from \cite{PhysRevLett.106.130402}. In this work, an entropy-based witness is introduced, constructed based on the Shannon entropies of the distributions $P^{B}_{x_0 \lvert \vec f}(q)$ and $\tilde{P}^{B}_{x_0 \lvert \vec f}(p)$:
\begin{align}
    &h(P\lvert x_{A}^{\vec f}=x_0) = -\int_{\mathbb{R}}P^{B}_{x_0 \lvert \vec f}(q) \log P^{B}_{x_0 \lvert \vec f}(q) dq,\\
    &h(\tilde{P}\lvert x_{A}^{\vec f}=x_0) = -\int_{\mathbb{R}}\tilde{P}^{B}_{x_0 \lvert \vec f}(p) \log \tilde{P}^{B}_{x_0 \lvert \vec f}(p) dp.
\end{align}
We can then define
\begin{align}
    &h^{B\lvert A}({\cal A}, \vec e^{\top}\vec {\hat x}) = \min_{\vec f \in \mathbb{R}^{2m}} \int_{\mathbb{R}} P_A(x_A^{\vec f}=x_0) h(P\lvert x_{A}^{\vec f}=x_0) dx_0, \\
    &h^{B\lvert A}({\cal A}, \vec e^{\top}\Omega\vec {\hat x}) = \min_{\vec f \in \mathbb{R}^{2m}} \int_{\mathbb{R}} P_A(x_A^{\vec f}=x_0) h(\tilde P\lvert x_{A}^{\vec f}=x_0) dx_0.
\end{align}
The original steering criterion that was proposed can be translated to our context as
\begin{equation}\label{eq:entropicIneq}
    h^{B\lvert A}({\cal A}, \vec e^{\top}\vec {\hat x}) + h^{B\lvert A}({\cal A}, \vec e^{\top}\Omega\vec {\hat x})  < \log (2 \pi e). 
\end{equation}
It is particularly useful to note that 
\begin{equation}\label{eq:VarEnt}
    \text{Var}_{\rm hom}^{B|A}\left(\mathcal{A},\frac{ \vec e^{\top} \Omega \vec{\hat x}}{2}\right)\text{Var}_{\rm hom}^{B|A}\left(\mathcal{A},\frac{ \vec e^{\top} \vec{\hat x}}{2}\right) \geqslant \frac{e^{2h^{B\lvert A}({\cal A}, \vec e^{\top}\vec {\hat x})}e^{2h^{B\lvert A}({\cal A}, \vec e^{\top}\Omega\vec {\hat x})}}{(2\pi e)^2}.
\end{equation}
When we combine this with the entropic inequality \eqref{eq:entropicIneq}, we can propose the steering witness 
\begin{equation}\label{eq:EntWitness}
    S_H({\cal A}) = \max_{\vec e \in \mathbb{R}^{2m'}} \left[2\pi e^{1-2h^{B\lvert A}({\cal A}, \vec e^{\top}\vec {\hat x})} -\frac{e^{2h^{B\lvert A}({\cal A}, \vec e^{\top}\Omega\vec {\hat x})-1}}{2\pi}\right]^+.
\end{equation}
In the limit for Gaussian states, we find that $S_H({\cal A}) = S^{hom}_R({\cal A})$. For more general states, we find that $S_H({\cal A}) \geqslant S^{hom}_R({\cal A})$. When comparing the metrological witness to the entropic one, we find a useful relation between the Fisher information and Shannon entropy in literature \cite{STAM1959101} that can be combined with Jensen's inequality to prove
\begin{equation}\label{eq:IneqEntPower1}
F^{B|A}_{\rm hom}\left(\mathcal{A},\frac{\vec e^{\top} \Omega \vec{\hat x}}{2}\right)  \geq 2 \pi e^{1-2h^{B\lvert A}({\cal A}, \vec e^{\top}\vec {\hat x})}.
\end{equation}
However, the variance and entropy power are also related to each, which was for example used to obtain \eqref{eq:VarEnt}. This leads to the inequality
\begin{equation}\label{eq:IneqEntPower2}
    \text{Var}_{\rm hom}^{B|A}\left(\mathcal{A},\frac{ \vec e^{\top} \Omega \vec{\hat x}}{2}\right) \geq \frac{e^{2h^{B\lvert A}({\cal A}, \vec e^{\top}\Omega\vec {\hat x})-1}}{2\pi}.
\end{equation}
When we combine both \eqref{eq:IneqEntPower1} and \eqref{eq:IneqEntPower2} we cannot establish a clear relation between the entropic witness $S_H({\cal A})$ and the metrological witness $S^{\rm hom}_{\rm max}(\mathcal{A})$. We explore which one of these two witnesses, based on the same homodyne measurement statistics, performs better for non-Gaussian states.

In Section \ref{sec:NonGaussianSteering}, we will explore the potential of the metrological protocol for an important class of two-mode non-Gaussian states, presented in Section~\ref{sec:PhotonSubtraction}, under ideal detection conditions. Details about the experimental estimation of these quantities for realistic detection schemes will be provided in Section \ref{sec:Realistic}.

\section{\label{sec:PhotonSubtraction} Mode-selective photon subtraction}
The protocol described in the previous Section is valid for any CV system, regardless of the nature of the state that we consider, as long as we have access to the marginals of the Wigner function along the desired axes in the phase space of each of the sub-systems. In this Section we will introduce the probe states that we shall consider throughout this paper, namely, photon-subtracted states. Different approaches can be followed to describe the generation of these states and obtain their Wigner function \cite{Tutorial2021, Walschaers2017_1,Walschaers2017_2,Braun2014}.

\begin{figure}[!htb]
\centering
\includegraphics[width=0.45\textwidth]{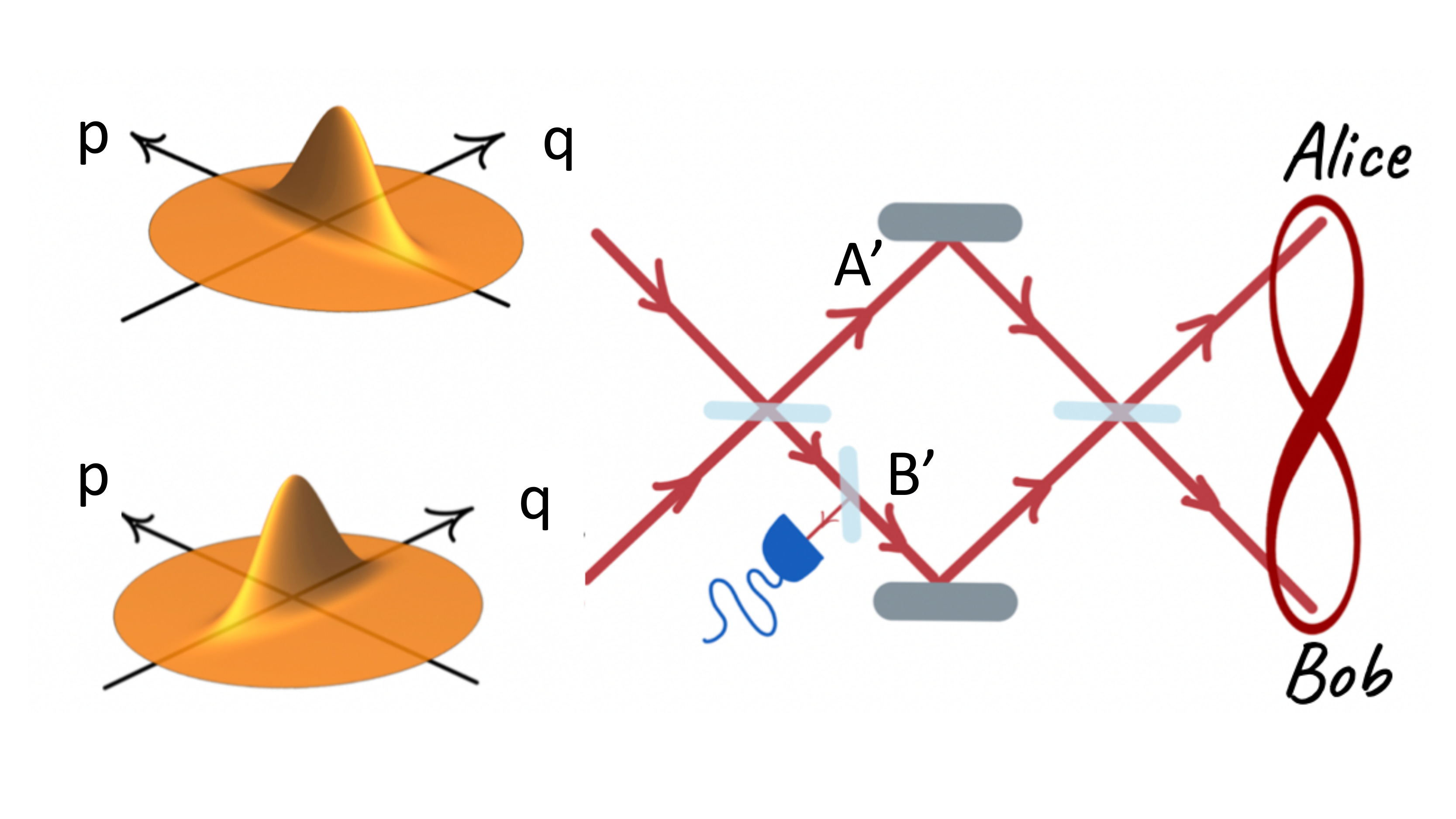}
\caption{Parameterized probe states: the non-Gaussian state obtained by subtracting one photon from one mode of a two-mode squeezed state, is passed through a beam splitter with a tunable transmissivity $\tau=\sin^{2}(\theta)$. }
\label{fig:PhotonSubt_DiffBasis}
\end{figure}

We focus on two-mode photon-subtracted states, where one mode is sent to Alice and the other to Bob. These states are generated through the setup sketched in Fig.~\ref{fig:PhotonSubt_DiffBasis}: two single-mode squeezed-vacuum states, squeezed in opposite quadratures, are mixed on a balanced beamsplitter to generate an EPR state. A single photon is subtracted in one of the two output modes, and the resulting state is mixed on a second beamsplitter with a variable reflectivity $\cos \theta$.

To accommodate losses and other experimental imperfections, we consider an arbitrary Gaussian two-mode state without mean field. We start by considering the state in the basis of EPR modes, which we denote $A'$ and $B'$, such that we have 
\begin{equation}
    W_G(\vec x) = \frac{e^{-\frac{1}{2}\vec x^{\top} V^{-1} \vec x}}{(2\pi)^2\sqrt{\det V}},
\end{equation}
where $V$ is the $4 \times 4$ covariance matrix of the state and $\vec x = \vec x_{A'} \oplus \vec x_{B'} = (x_{A'},p_{A '}, x_{B'}, p_{B'})^{\top}$ contains the coordinates in phase space. Subsequently, we subtract a photon in the first mode $A'$, such that the relevant Wigner function is given by \cite{Tutorial2021}
\begin{equation}\label{eq:Wigner_Photon_Subtracted}
\begin{split}
    W^{-}(\vec x)=\frac{\norm{P_{A'}(\1-V^{-1}) \vec x}^2 - \tr(P_{A'} V^{-1}) + 2 }{\tr(V_{A'}-\1)} W_G(\vec x),
\end{split}
\end{equation}
where $P_{A'}$ is a projector on the first mode, given by
\begin{equation}
    P_{A'} = \begin{pmatrix} 1 & 0 & 0& 0\\
    0& 1& 0&0\\
    0&0&0&0\\
    0&0&0&0
    \end{pmatrix},
\end{equation}
and $V_{A'}$ is the covariance matrix for the reduced state of the first mode, given by $V_{A'} = P_{A'}V P_{A'}$.\\

In the ideal setting of Fig.~\ref{fig:PhotonSubt_DiffBasis}, we can describe the covariance matrix as
\begin{equation}\label{eq:Vmat}
V= \frac{1}{2}\left(\begin{matrix}
r_1+\frac{1}{r_2} && 0 && \frac{1}{r_2}-r_1 && 0 \\
0 && r_2+\frac{1}{r_1} && 0 && r_2-\frac{1}{r_1}\\
\frac{1}{r_2}-r_1 && 0 && r_1+\frac{1}{r_2} && 0\\
0 && r_2-\frac{1}{r_1} && 0 && r_2 + \frac{1}{r_1}\\
\end{matrix}\right),
\end{equation}
where $r_{i}=10^{\frac{s_{i}}{10}}$, with $s_i$ representing the squeezing parameter of the squeezed mode $i=1,2$, given in decibels (dB), and the squeezing is applied in opposite quadratures.

Photon losses can be described in an open quantum system approach, as an interaction of the system with the environment \cite{Walschaers2019}. When the losses are the same in both modes, the effect can be entirely absorbed within the covariance matrix, regardless of whether they act before or after the photon subtraction. The effect of losses can then be modeled by modifying the covariance matrix in the following way
\begin{equation}\label{eq:losses}
    V \mapsto (1-\eta) V+\eta \1,
\end{equation}
where $\eta\in \left[0,1\right]$ represents the amount of loses.

We apply a tuneable beamsplitter after the local photon subtraction. The parameter $\theta,$ that parameterizes the non-Gaussian states, determines the transmissivity of the beamsplitter ($T=\sin^{2}(\theta)\in\left[0,1\right]$), whose effect on the quadratures of the phase space is described by the matrix
\begin{equation}
M(\theta)=\left(\begin{matrix}
\cos(\theta) && 0 && \sin(\theta) && 0\\
0 && \cos(\theta) && 0 && \sin(\theta)\\
-\sin(\theta) && 0 && \cos(\theta) && 0\\
0 && -\sin(\theta) && 0 && \cos(\theta)\\
\end{matrix}\right).
\end{equation}
The Wigner function of the resulting state which is sent to Alice and Bob is then written as 
\begin{equation}\label{eq:wigFinal}
W^{-}_{\theta}(\vec{x_A}\oplus\vec{x_B})=W^{-}(M(\theta)^{T} \vec{x}).
\end{equation}
The set of non-Gaussian probe states include $\theta=0$ and $\theta=\pi/2$, \ie zero transmissivity and zero reflectivity, which leave the state untouched (up to a swap of the modes). In the former cases, the photon is subtracted in Alice's mode, whereas in the latter case it is subtracted in Bob's mode. Here, we expect an enhancement of Gaussian quantum correlations of the EPR state through the generation of non-Gaussian features. On the other hand, $\theta=\pi/4$ would undo the correlations in the absence of photon subtraction. However, if a photon is subtracted, the second beamsplitter delocalises the non-Gaussian features of the state over Alice's and Bob's modes. In this case, we witness a purely non-Gaussian quantum correlation, exclusively generated after photon subtraction, as no correlation is encoded in the covariance matrix of the corresponding state. This implies that Gaussian protocols like those based on Reid's criteria are expected to fail to witness steering.

\section{\label{sec:NonGaussianSteering} Ideal detection of non-Gaussian quantum steering}
In this Section we consider the protocol established in Section \ref{sec:Fisher} for detection of steering using as probe states the photon subtracted states introduced in Section \ref{sec:PhotonSubtraction}. We first consider ideal results, neglecting the effect of any losses in the system. After that, we study the effect of losses in each possible scenario in an analytical way.

\subsection{\label{sec:GaussianSteering} Gaussian witnesses for quantum steering}
Before considering the non-Gaussian scenario, with the double purpose of validating the protocol and setting up comparison for the fore-coming results, we analyse steering in Gaussian two mode-squeezed states (\ie before photon subtraction in Fig.~\ref{fig:PhotonSubt_DiffBasis}).

Because Alice and Bob only control a single mode, we can simplify our notation compared to Section \ref{sec:ProtocolCV}, by naming the measured quadratures on Bob's side 
\begin{align}
    &\hat q \coloneqq  \vec e^{\top}\vec{\hat x},\label{eq:qishere}\\
    &\hat p \coloneqq \vec e^{\top} \Omega \vec{\hat x} \;.\label{eq:pishere}
\end{align} 
Alice's choice of a phase space axis is equivalent to choosing an angle $\phi$ such that she measures 
\begin{equation}\label{eq:Alice}
\hat x_A (\phi) \coloneqq \cos \phi \,\hat q_A + \sin \phi \,\hat p_A \;,\end{equation}
which means that $\hat x_A (\phi)$ is any quadrature in Alice's mode.
\\

\begin{figure}[!htb]
\centering
\includegraphics[width=0.45\textwidth]{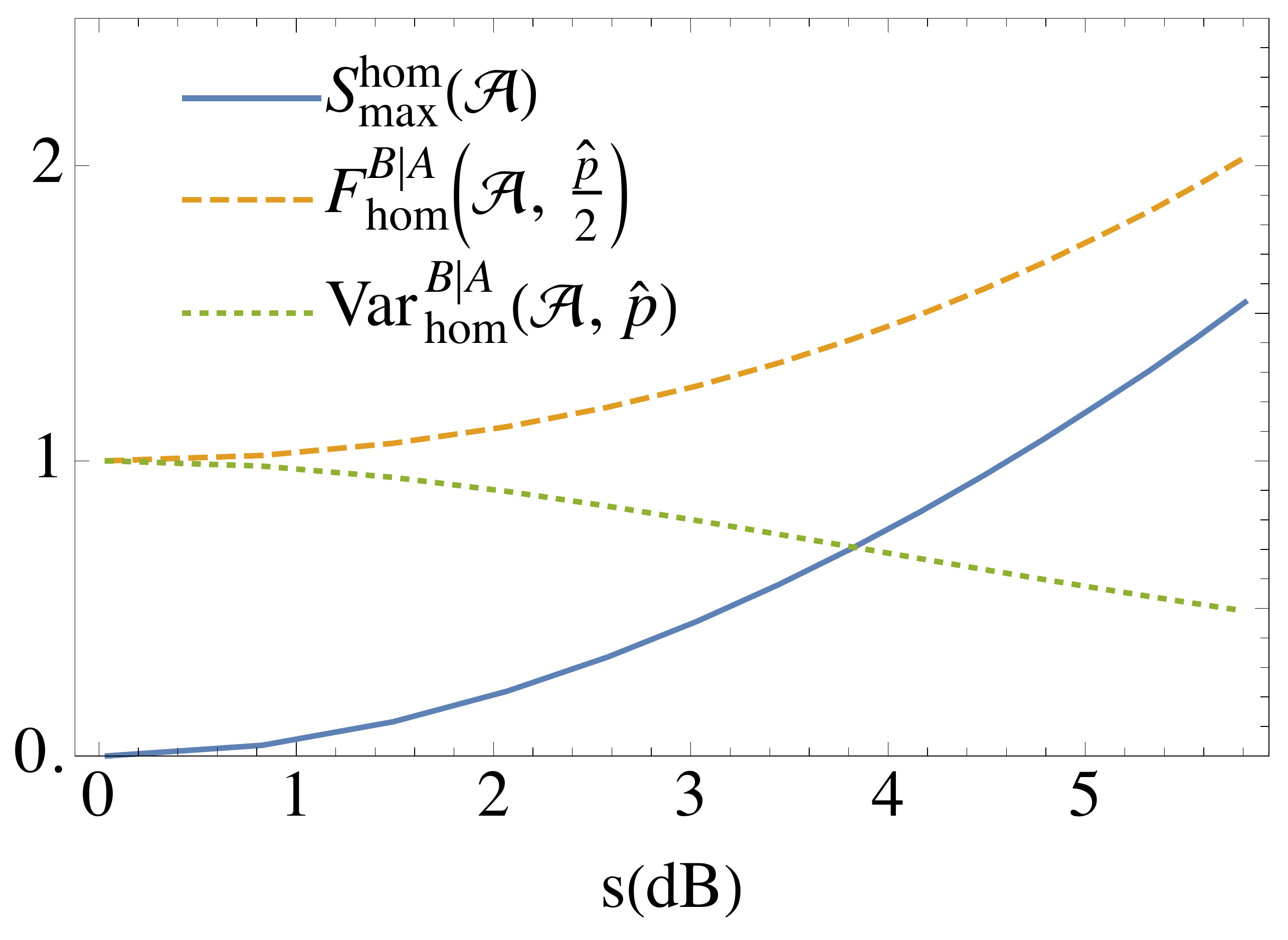}
\caption{Witnessing steering in two-mode squeezed states with equal squeezing in both modes as a function of the squeezing level $s$. We show the results using the steering witness ~(\ref{eq:Steering_witness_hom}), which is in this case identical to Reid's criterion $S_{R}^{\rm hom}(\A)$ (recall that any value larger than zero implies quantum steering). We choose Alice's measurement settings \eqref{eq:Alice} as $\phi = 0$ and $\phi=\pi/2$ to achieve a maximal value of the steering witness (see main text). An optimization was performed over all possible choices of the generator of displacements on Bob's side.}
\label{fig:witnessing_steering_Gauss_states}
\end{figure}

\begin{figure}[!htb]
\centering
\includegraphics[width=0.45\textwidth]{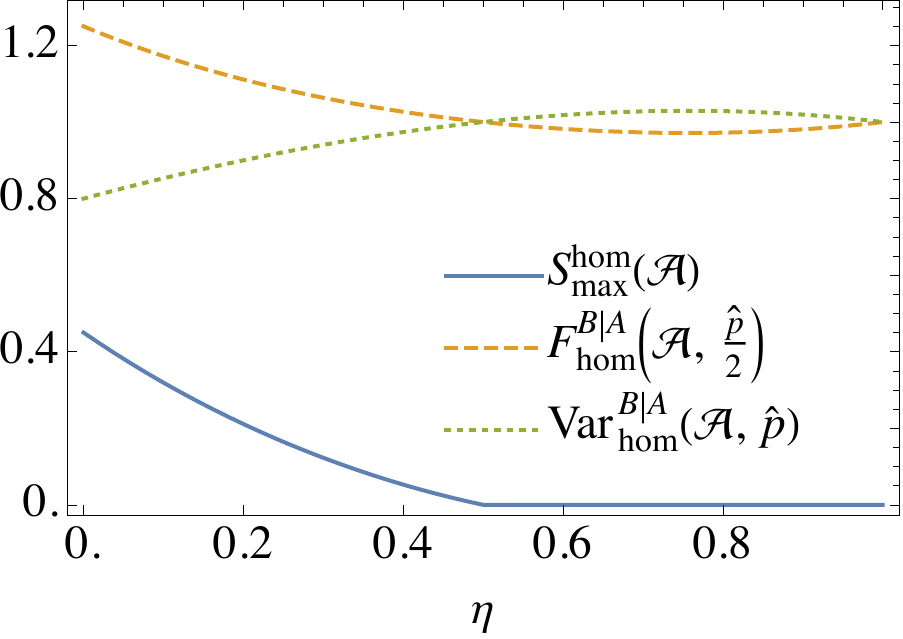}
\caption{Effect of losses in the steering witness \eqref{eq:Steering_witness_hom}, for a two-mode squeezed state, as considered in Fig.~\ref{fig:witnessing_steering_Gauss_states} for a level of squeezing of $s=3{\rm dB}$.} 
\label{fig:losses_steering_Gauss_states}
\end{figure}

In Fig.~\ref{fig:witnessing_steering_Gauss_states}, we present the results obtained when we consider an EPR state by setting equal squeezing values, i.e., $s\equiv s_1=s_2$, in \eqref{eq:Vmat}. We analyse the violation of the metrological inequality after homodyne detection by Alice, considering the largest possible violation obtained over all possible choices of the quadrature on Bob's side over which the displacement takes place, as prescribed by the maximization in \eqref{eq:Steering_witness_hom}. 
As there is no global phase dependence in the EPR state, Alice is completely free to choose one measurement setting $\phi$. Bob will thus have to choose $\vec e$ such that $\hat q$ is maximally correlated with $\hat x_A (\phi)$. This choice immediately fixes the second quadrature $\hat p$ that Bob will measure through \eqref{eq:pishere}. The largest value of the steering witness \eqref{eq:Steering_witness_hom} will then be obtained if Alice chooses a second measurement setting that measures the quadrature that is most strongly correlated with $\hat p$. A key property of the state is that the correlation between Alice and Bob's measurements is the strongest when they measure the same quadrature (\ie when their homodyne measurements are in phase), which means that Alice's second setting should be set to $\phi + \pi/2$.\\
In Fig.~\ref{fig:losses_steering_Gauss_states}, we show the effect of photon losses \eqref{eq:losses} for the same type of states as in Fig.~\ref{fig:witnessing_steering_Gauss_states}, for $3 {\rm dB}$ squeezing. The latter is relevant to further understand the relation between Gaussian and non-Gaussian steering and the fundamental differences that can arise between one and the other.\\

\subsection{ Quantum steering after local photon subtraction}
Moving now to the non-Gaussian realm, the natural first scenario to consider is the subtraction of one photon in one of the two correlated modes $A'$ or $B'$, in the previously considered Gaussian scenario. This corresponds to $\theta=0$ or $\theta=\pi/2$ in the tuneable beam splitter in Fig.~\ref{fig:PhotonSubt_DiffBasis}. 
A recent result shows that Gaussian steering before photon subtraction is a sufficient condition for remotely generating Wigner negativity \cite{Walschaers2020}. In the following we explore a complementary property and investigate how local photon subtraction affects the steering of the state.

\begin{figure}[!htb]
\centering
\includegraphics[width=0.35\textwidth]{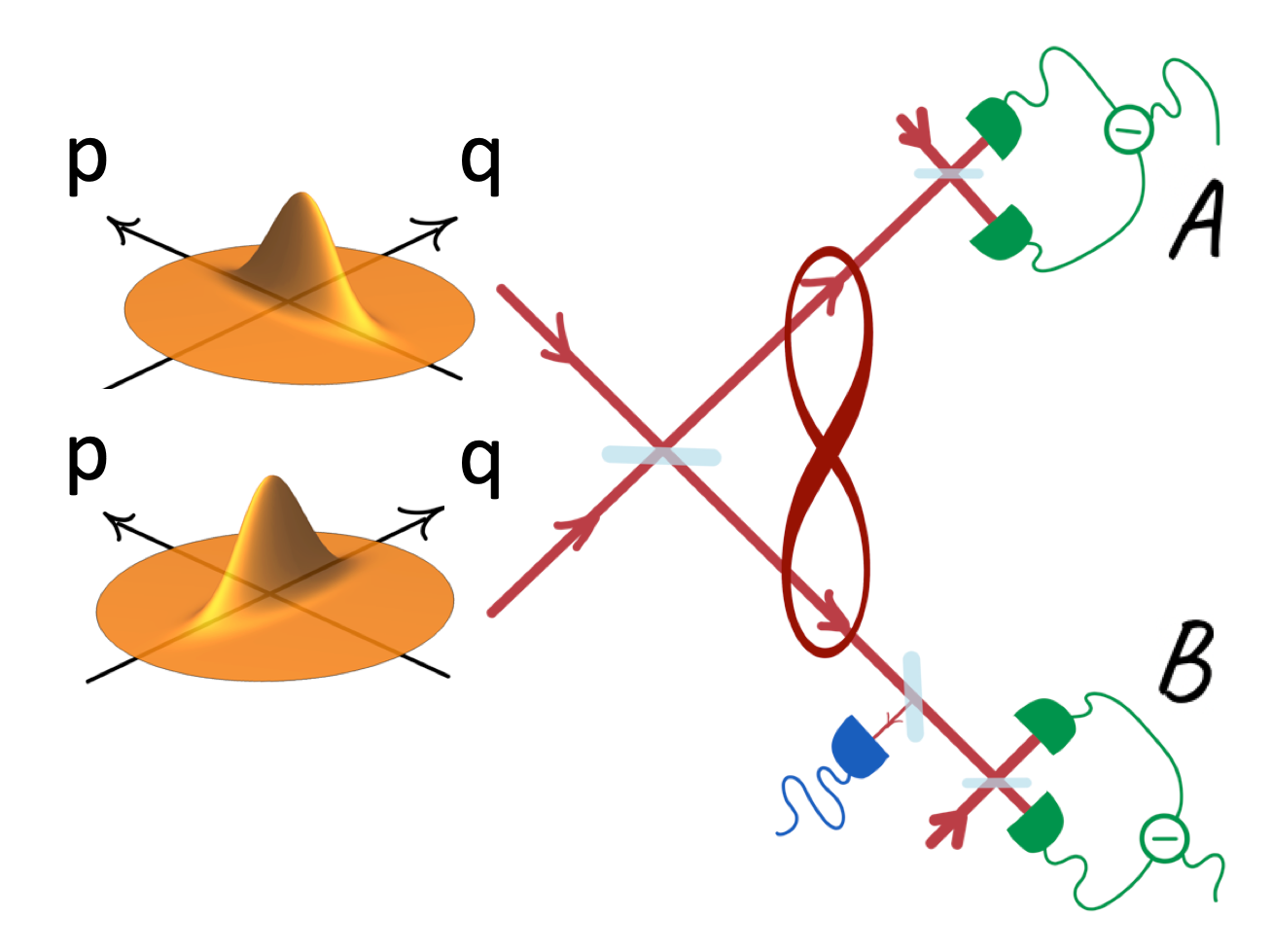}\\
\includegraphics[width=0.46\textwidth]{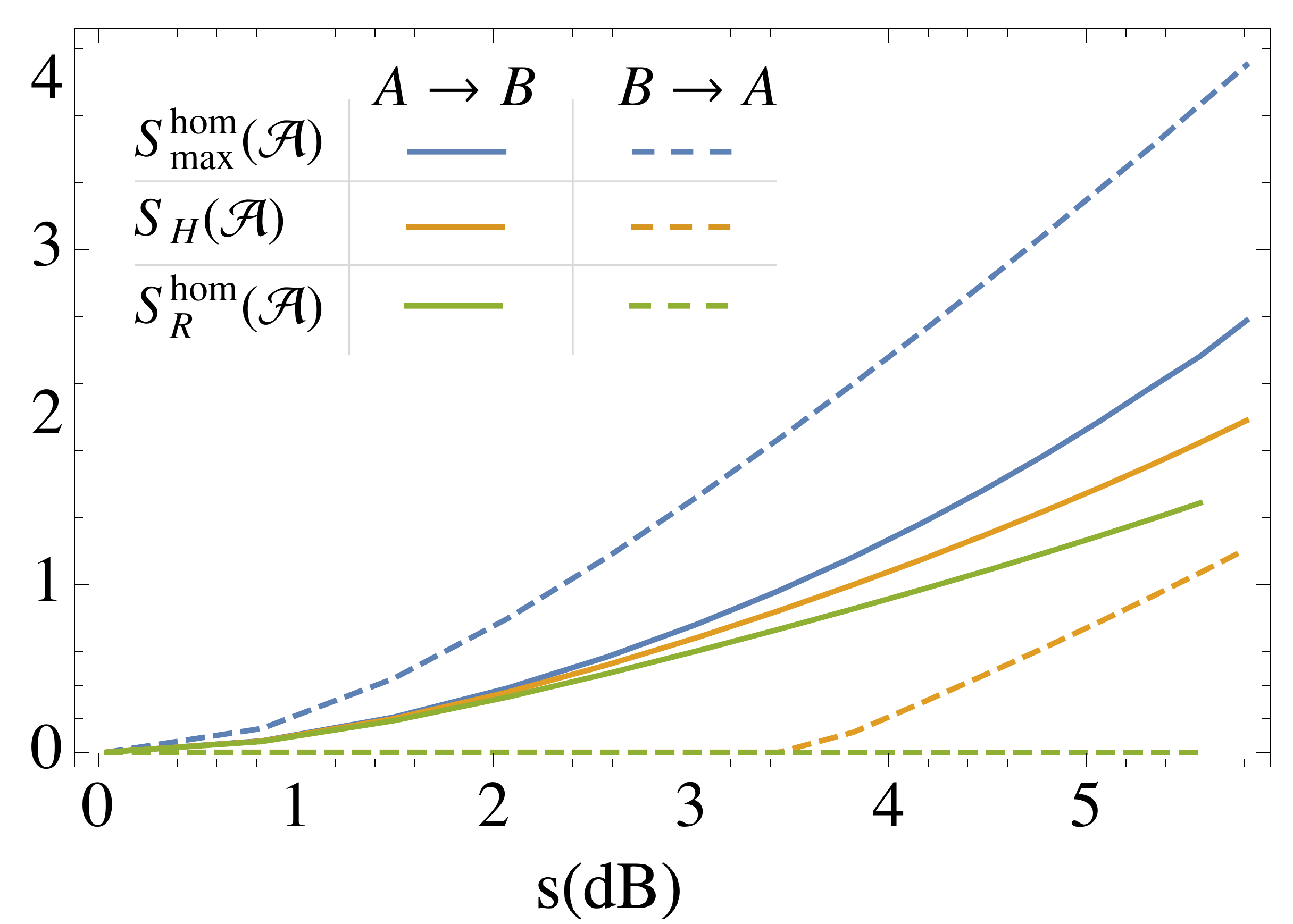}
\caption{Steering in photon-subtracted states corresponding to the choices of $\theta=0$ and $\theta=\pi/2$ in the tuneable beamsplitter in Fig.~\ref{fig:PhotonSubt_DiffBasis}. Solid curves correspond to $A$ steering $B$, whereas dashed lines correspond to the scenario where $B$ is steering $A$. In green we show the observations arising from the application of Reid's criterion $S_{R}^{\rm hom}(\A)$ \eqref{eq:ReidHom}. In this case, no steering from the mode where the photon was subtracted can be observed. In blue, we plot the steering witness $S_{\text{max}}^{\text{hom}}(\A)$ \eqref{eq:Steering_witness_hom} for the same set of states and we can see that a violation of the inequality is attained in the direction where no Gaussian EPR steering is observed through Reid's criteria, and remarkably, this violation is larger than in the opposite direction. This represents a remarkable signature of non-Gaussian steering. In orange we finally show that the entropic witness can pick up on steering from $B$ to $A$ only when there is a sufficient amount of steering. On the one hand, this clearly highlights the capabilities of the entropic witness to detect non-Gaussian steering. On the other hand, it also shows that the metrological witness can detect steering in parameter regimes where the entropic witness cannot. We also observe that in both directions $S_{R}^{\rm hom}(\A)\leq S_{H}(\A) \leq S_{\text{max}}^{\text{hom}}(\A)$.}
\label{fig:witnessing_steering_Photon_subtracted_Corr_Basis}
\end{figure}

States obtained by local photon subtraction are non-symmetric. Wigner negativity, for example, is only present in the reduced state of the mode complementary to the one where the photon was subtracted. However, the Wigner negativity of the two-mode Wigner function is larger than the single-mode Wigner negativity \cite{xiang2021quantification}, which indicates the presence of non-local effects.  In the same way, one would expect that steering, which is intrinsically a one-sided property, should not behave in the same way in both directions, \ie steering from the mode where the photon was subtracted to the complementary mode is expected to be different from the steering in the opposite direction. To check this, in Fig.~\ref{fig:witnessing_steering_Photon_subtracted_Corr_Basis} we show the steering witness, as measured in the two directions. In the green curves, we use Reid's criterion \eqref{eq:ReidHom}, which leads to strongly asymmetric results, as no EPR steering from the mode where the photon was subtracted is observed. Yet, remarkably, the metrological witness \eqref{eq:Steering_witness_hom} not only witnesses steering from the photon subtracted mode, but actually leads to a larger value for the steering witness. This observation contrasts with what one would expect from Reid's criterion, thus clearly showing new non-Gaussian behaviour. In a more operational sense, this result shows that non-Gaussian steering from the photon subtracted mode to the complementary mode can considerably enhance the inference of displacements in the complementary mode. The entropic witness \eqref{eq:EntWitness} is also shown to detect steering from the photon subtracted mode, but only when there is sufficient squeezing in the initial squeezed modes. This means that there is non-Gaussian steering that can be detected by the metrological witness, but not by the entropic one. Furthermore, we observe that the metrological witness systematically produces larger values than the entropic one (both coincide for Gaussian steering).

\begin{figure}[!htb]
\centering
\includegraphics[width=0.45\textwidth]{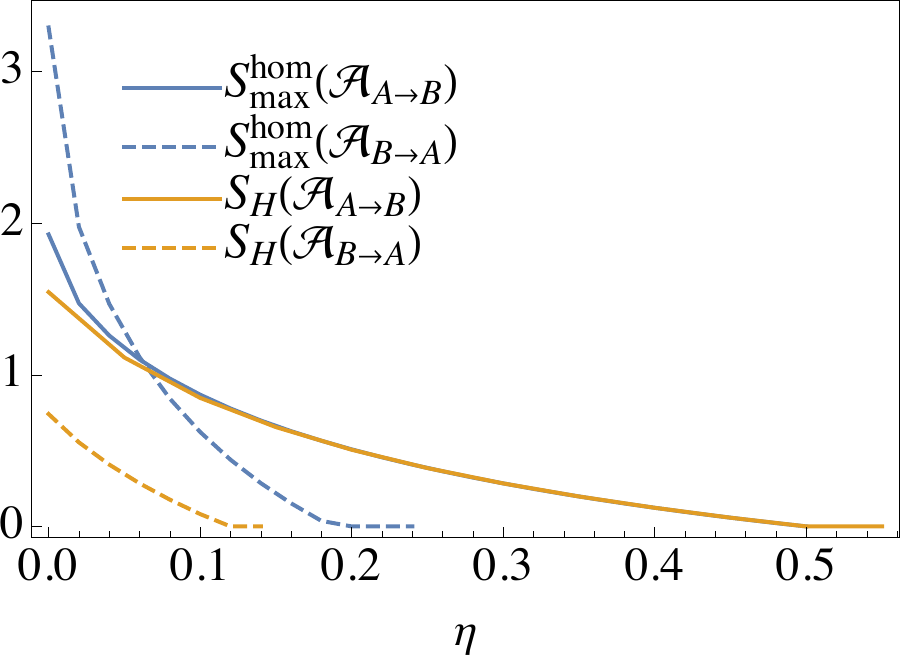}
\caption{Steering in photon-subtracted state, corresponding to the choices of $\theta=0$ and $\theta=\pi/2$ in Fig.~\ref{fig:PhotonSubt_DiffBasis}, after uniform loss $\eta$ \eqref{eq:losses}. In particular we consider the photon subtraction in a $5$dB squeezed state. In accordance with Fig.~\ref{fig:witnessing_steering_Photon_subtracted_Corr_Basis}, we can observe that in the absence of losses the violation of inequality \eqref{eq:Metrological_Inequality} is larger when we consider the steering from the photon-subtracted mode than in the other direction. Yet, with increasing losses, the former decreases much faster (vanishing at $18\%$ losses) than the steering from the complementary mode (vanishing at $50 \%$ losses, as in Fig.~\ref{fig:losses_steering_Gauss_states}). When comparing the metrological witness $S_{\rm max}^{\rm hom}(\cal A)$ to the entropic witness $S_H(\A)$ we observe that for the steering from $A$ to $B$ (where the steering resembles Gaussian steering) for the larger values of $\eta$ both witnesses coincide. However, the metrological witness clearly outperforms the entropic one. We even find parameter ranges where the metrological witness detects steering that goes undetected by the entropic witness.} 
\label{fig:losses_steering_Photon_subtracted_Corr_Basis}
\end{figure}

In Fig.~\ref{fig:losses_steering_Photon_subtracted_Corr_Basis}, we consider the effect of losses as we did previously for Gaussian states. The goal is to understand how resilient the witnesses are and how they are connected to the Gaussian scenario. As discussed in Section ~\ref{sec:PhotonSubtraction}, uniform losses in photon-subtracted states can be modeled by modifying the initial Gaussian covariance matrix as if the losses occurred at this initial stage. In other words, we analyze how photon subtraction affects Fig.~\ref{fig:losses_steering_Gauss_states}, with the remark that steering is not symmetric, as we already discussed in Fig.~\ref{fig:witnessing_steering_Photon_subtracted_Corr_Basis}.

There are some remarkable features observed in Fig.~\ref{fig:losses_steering_Photon_subtracted_Corr_Basis}. In the first place, as we previously observed in Fig.~\ref{fig:witnessing_steering_Photon_subtracted_Corr_Basis} in the absence of losses, steering from the photon-subtracted mode seems to be stronger than the steering from the complementary mode, in the sense that a larger violation of inequality
~\eqref{eq:Metrological_Inequality} is attained. Nevertheless, when we consider the effect of losses, we observe a much faster decay in the former which renders it harder to witness in a real experiment. For the entropic witness \eqref{eq:EntWitness} we see a somewhat slower decay. However, given that the initial value of the witness in absence of losses is much smaller than for the metrological witness, we still find that the entropic witness is less tolerant to losses. On the other hand, regardless of the witness we use, steering from the complementary mode goes away for the same amount of losses as the Gaussian steering does. These observations, together with the impossibility of witnessing the steering from the photon subtracted mode using Reid's criterion, lead us to interpret the steering from the complementary mode as an enhanced type of Gaussian steering, while the steering from the photon subtracted mode appears to be purely non-Gaussian, stronger, but less resilient to losses. This behavior is equivalent for other values of squeezing, and the amount of losses required to destroy the steering from the photon subtracted state increases with it.\\

\subsection{Purely non-Gaussian quantum steering}
\begin{figure}[!htb]
\centering
\includegraphics[width=0.46\textwidth]{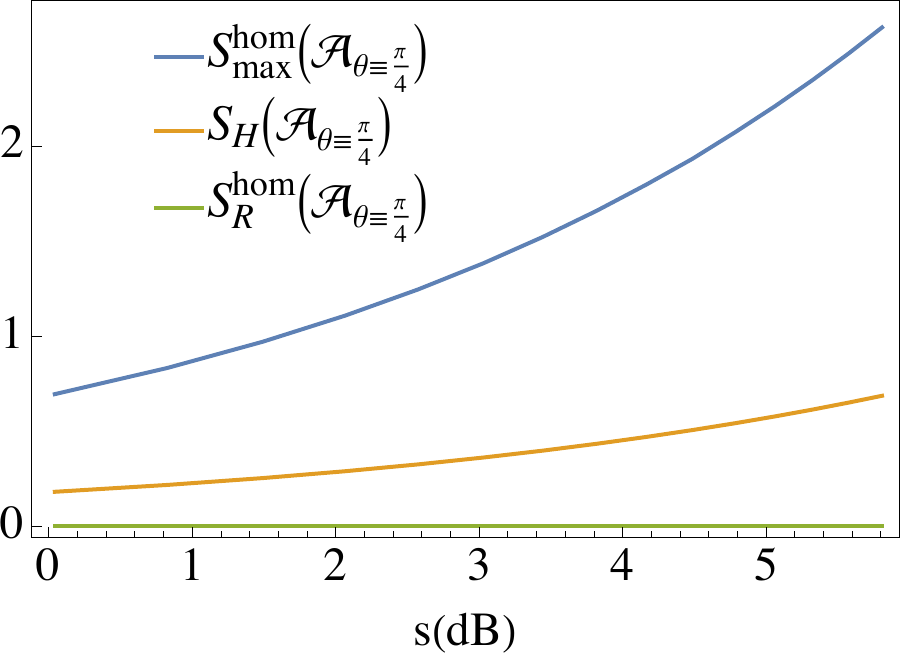}
\caption{Witnessing non Gaussian steering,  for the state generated by having $\theta=\pi/4$ in the second beam splitter in Fig.~\ref{fig:PhotonSubt_DiffBasis}.  In green we show the absence of violation of Reid's criterion, expressed in the form of witness \eqref{eq:ReidHom}. In blue and orange we observe how an increasing violation of the metrological and entropic inequalities, respectively, are obtained as the squeezing level increases, starting from a finite value of the witnesses \eqref{eq:Steering_witness_hom} and \eqref{eq:EntWitness} for arbitrarily low squeezing.} 
\label{fig:witnessing_steering_Photon_subtracted_UnCorr_Basis}
\end{figure}

The most striking shortcomings of considering Gaussian measurements of steering do naturally arise when we consider purely non-Gaussian correlations. In the present section we analyze the steering in the state obtained after setting $\theta=\pi/4$ in the second beam splitter in Fig.~\ref{fig:PhotonSubt_DiffBasis}.
The final state is equivalent to the state that would be obtained by subtracting a single photon from a superposition of the two initially uncorrelated squeezed modes, which is a non-local non-Gaussian operation.  The non-Gaussian nature of these correlations can be seen in the Wigner function \eqref{eq:Wigner_Photon_Subtracted}, whose Gaussian part factorizes for $\theta=\pi/4$. 
In Fig.~\ref{fig:witnessing_steering_Photon_subtracted_UnCorr_Basis}, we show the analysis of the steering as a function of the squeezing level for this scenario. We consider, as before, two equally squeezed modes, squeezed in opposite quadratures, \ie, setting $r_1=r_2$ in \eqref{eq:Vmat}. We show how Reid's criterion fails to witness any quantum steering in this case, while we witness steering through the witnesses \eqref{eq:Steering_witness_hom} and \eqref{eq:EntWitness}. Even for arbitrarily low amounts of squeezing we find that these witnesses do not tend to zero, which is fundamentally different to the scenario obtained after local photon subtraction. Thus, we observe that by means of a non-local non-Gaussian operation a finite amount of steering is created for arbitrarily low squeezing. This observation is in agreement with those obtained when measuring entanglement in this kind of non-Gaussian states \cite{Tutorial2021}, and can intuitively be understood in the following way: For an arbitrarily low amount of squeezing both modes are to good a approximation a superposition of vacuum and two-photon Fock states. After photon subtraction in a balanced superposition of the two, we obtain an entangled two-mode state, given by $(|01\rangle+|10\rangle)/\sqrt{2}$, which is a Bell state.\\

\begin{figure}[!htb]
\centering
\includegraphics[width=0.46\textwidth]{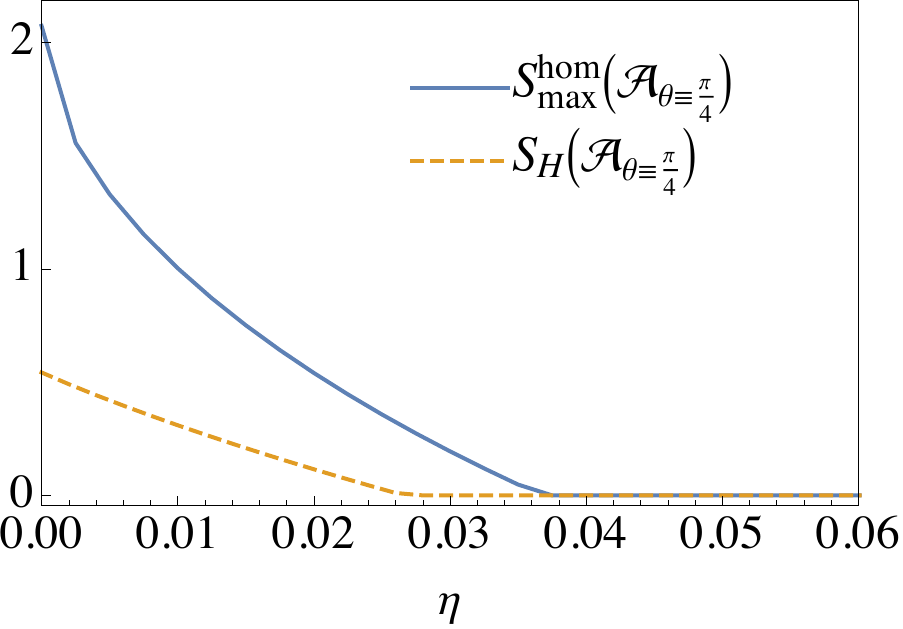}
\caption{Effect of photon losses on purely non-Gaussian steering, quantified by both the metrological witness (blue solid curve) and the entropic witness (dashed orange curve) for the state generated with $\theta=\pi/4$ in the second beamsplitter in Fig.~\ref{fig:PhotonSubt_DiffBasis}.} 
\label{fig:losses_steering_Photon_subtracted_UnCorr_Basis_WitnessComparison}
\end{figure}

In this case of purely non-Gaussian steering, both the metrological and the entropic witness have been shown to be effective. However, in Fig.~\ref{fig:losses_steering_Photon_subtracted_UnCorr_Basis_WitnessComparison} we explore how both witnesses behave in the presence of losses. As for Fig.~\ref{fig:losses_steering_Photon_subtracted_Corr_Basis}, we once again find that the metrological witness is more resilient to losses. Similar plots can be produced for all squeezing levels, showing the same behaviour.

Summing up all our comparisons between the entropic and metrological witnesses, we conclude that there are cases where the metrological witness can detect steering that goes undetected by the entropic witness. We have not found any opposite case, leading us to suggest that for single-photon subtracted states the metrological witness tends to outperform the entropic one. Therefore, we will focus our attention on the metrological witness in the remainder of this article.

\begin{figure}[!htb]
\centering
\includegraphics[width=0.46\textwidth]{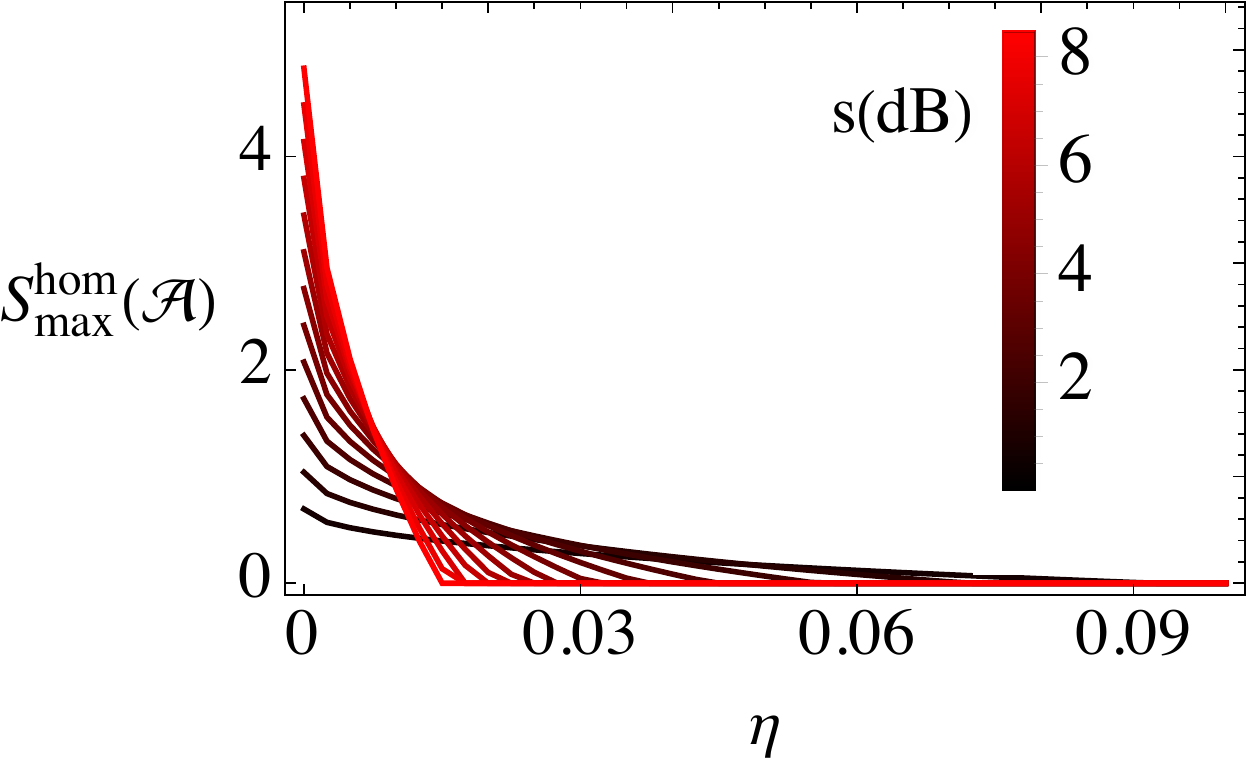}
\caption{Effect of photon losses on purely non-Gaussian steering for the state generated with $\theta=\pi/4$ in the second beamsplitter in Fig.~\ref{fig:PhotonSubt_DiffBasis}. We observe a reduction in the resilience to noise as the level of squeezing in the initial states increase. The latter might be linked to the fact that the effect of losses is more severe in single mode squeezed states the larger the squeezing is \cite{Bachor2019}.   } 
\label{fig:losses_steering_Photon_subtracted_UnCorr_Basis}
\end{figure}

In Fig.~\ref{fig:losses_steering_Photon_subtracted_UnCorr_Basis}, we show how the witness \eqref{eq:Steering_witness_hom} behaves for these states under the effect of uniform photon losses. The behavior is very different to what we observe in the Gaussian scenario. First, we observe a very weak resilience to noise compared to the former one.  Yet, the most striking feature is that this resilience decreases as the level of squeezing (and thus steering) increases, contrary to what happens in the correlated basis ($\theta = 0$), even for the steering from the mode in which the photon was subtracted. \\

\begin{figure}[!htb]
\centering
\includegraphics[width=0.46\textwidth]{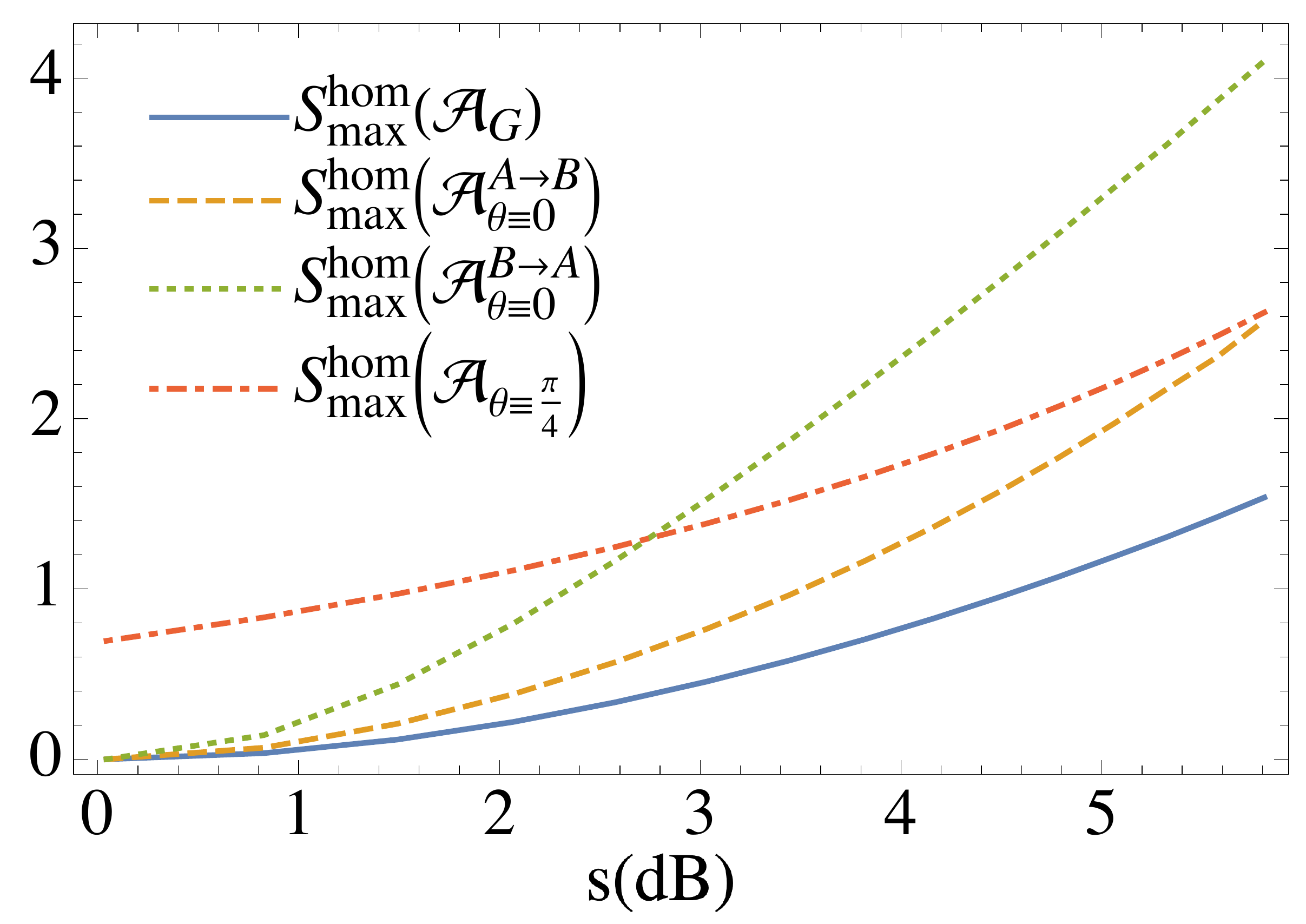}
\caption{Comparison of the behavior of the steering witness in the different scenarios considered so far.  $\A_{G}$ stands for the assemblage corresponding to the Gaussian state, $\A_{\theta=\{0,\pi/4\}}$ stand for the assemblages corresponding to the states generated through the choices of $\theta=\{0,\pi/4\}$ in the tuneable beam splitter in Fig.~\ref{fig:PhotonSubt_DiffBasis}. $\A_{\theta=0}^{A\rightarrow B(B\rightarrow A)}$ stand for the two non equivalent directions in which steering can occur in the case $\theta=0$, the convention is the same as in Fig.~\ref{fig:witnessing_steering_Photon_subtracted_Corr_Basis}.    } 
\label{fig:witnessing_steering_Comparison}
\end{figure}

Finally, in Fig.~\ref{fig:witnessing_steering_Comparison}, we show a comparison of how the witness \eqref{eq:Steering_witness_hom} behaves in the different scenarios that we have considered, namely, the Gaussian case and the photon subtracted states obtained by the procedure described in Fig.~\ref{fig:PhotonSubt_DiffBasis}, for $\theta=0$, considering both steering from Alice to Bob and from Bob to Alice, and for the purely non-Gaussian case $\theta=\pi/4$.  

\section{\label{sec:Realistic} Realistic detection of non-Gaussian quantum steering}
The approach followed so far considers the ideal scenario in which we can condition the state in Bob's steered mode on a definite outcome of Alice's measure. Yet, clearly, the latter is equivalent, from an experimental point of view, to have access to an infinite amount of data, namely, to sample the whole continuum of possible outcomes. In this Section we approach the problem in a more realistic fashion, by discretizing the set of Alice's measurement outcomes, in a way that we no longer condition Bob's state on a single outcome but rather in a mixture of the conditioned states belonging to a given \textit{bin} on Alice's side. 

First, we will analyze this scenario in an analytic way, to understand the limitations of this procedure. Later, keeping in mind the results from the former analysis, we consider the more realistic scenario, in which we study the protocol by simulating homodyne detection with rejection sampling.

\subsection{Conditioning on finite data}\label{sec:Disc}
Following the previous idea, the measurement outcomes that Alice communicates are rather determined by a histogram than by a probability density of continuous quadrature measurement outcomes. Therefore, we partition the real line corresponding to the outcomes of the quadrature measured by Alice in a series of $n$ bins
\begin{equation}
{\cal I} = \{{\cal I}_{1},\dots, {\cal I}_n \} \;,
\end{equation}
such that $\Re = \bigcup_k {\cal I}_{k}$ and we note ${\cal I}_k = [l_{k-1},l_{k})$, with $l_0 = -\infty$ and $l_n = \infty$. The probability of measurement outcomes now is described by
\begin{equation}
P_A[x_A(\phi)\in \mathcal{I}_{k}]=\int_{l_{k-1}}^{l_{k}} P_A[x_A(\phi)=x_{0}] dx_0 \;,
\end{equation}
where $x_A(\phi)$ stands for the quadrature measured by Alice, for which we will keep the notation introduced in Section \ref{sec:NonGaussianSteering} for two-mode states. An assemblage then gives a discrete sum of the form
\begin{equation}
\mathcal{A}(\mathcal{I}_{k},\hat x_A(\phi))=P_A[x_{A}(\phi)\in \mathcal{I}_{k}] \hat{\rho}^{B}_{\mathcal{I}_{k} | \phi},
\end{equation}
where
\begin{equation}\hat{ \rho}^{B}_{\mathcal{I}_{k}|\phi}=\int_{l_{k-1}}^{l_{k}} P_A[x_A(\phi)=x_{0}] \hat{ \rho}^{B}_{x_0|\phi}dx_0\label{eq:IntervalCondState}\end{equation}
is the conditional state on Bob's side after the measurement by Alice of quadrature $x_{A}(\phi)$ falls in the bin $\mathcal{I}_k$. This state is a mixture of all the conditional states conditioned on definite quadrature outcomes, with a weight given by the marginal probability density $P[x_A(\phi)]$.

The conditional 
FI now has to be calculated considering the discrete assemblage
\begin{equation}
 F^{B|A}_{\rm disc}(\mathcal{A},\hat H)=\max_{\phi\in\left[0,2 \pi\right)}\sum_{k}P_A[x_A(\phi)\in \mathcal{I}_{k}] F^{B}_{\xi}[P^{B}_{\mathcal{I}_{k}|\phi}] \;,
\end{equation}
where $F^{B}_{\xi}[P^{B}_{\mathcal{I}_{k}|\phi}]$ is computed according to \eqref{eq:Classical_FI},  with $P^{B}_{\mathcal{I}_{k}|\phi}(x|\xi)$ being the marginal along the displaced quadrature, characterized by $\vec e$, conditioned on the displacement $\xi$. This probability density can be obtained as $P^{B}_{\mathcal{I}_{k},\phi}(x|\xi)=\int_{l_{k-1}}^{l_{k}} P_A[x_A(\phi)=x_{0}] P^{B}_{x_0|\phi}(x|\xi)dx_0$.  Due to the convexity of the FI we find
\begin{equation}
F^{B|A}_{\rm disc}(\mathcal{A},\hat H)\leq F^{B|A}_{\rm hom}(\mathcal{A},\hat H) \;,
\end{equation}
where $F^{B|A}_{\rm hom}(\mathcal{A},\hat H)$ is the conditional FI when no coarse-graining is considered.

If we consider the estimation of displacements $\xi$ along the position quadrature $\hat{q}$ \eqref{eq:qishere}, generated by the Hamiltonian $\hat H=\hat{p}/2$, using \eqref{eq:pishere},
\begin{equation}
\begin{split}
  F^{B|A}_{\rm disc}\left(\mathcal{A},\frac{\hat{p}}{2}\right)=\max_{\phi\in\left[0,2 \pi\right)}\sum_{k}&P_A[x_A(\phi)\in \mathcal{I}_{k}]\int_{\Re}P^{B}_{\mathcal{I}_k|\phi}(q-\xi)\\ &\times \left\{ \frac{\partial \log\left[P^{B}_{\mathcal{I}_k|\phi}(q-\xi)\right]}{\partial\xi}\right\}^2 dq \;,
 \end{split}
\end{equation}
where  we made use of the identity $P^{B}_{\mathcal{I}_k|\phi}(q|\xi) = P^{B}_{\mathcal{I}_k|\phi}(q-\xi)$. 

The conditional variance of the generator $\hat p/2$ is calculated in a similar way 
\begin{equation}\label{eq:conditional_Quantum_Variance_Bined}
 \text{Var}^{B|A}_{\rm disc}\left(\mathcal{A},\frac{\hat{p}}{2}\right)=\min_{\phi\in\left[0,2 \pi\right)}\sum_k P_A[x_A(\phi)\in \mathcal{I}_k] \text{Var}\left(\hat \rho^{B}_{\mathcal{I}_{k}|\phi},\frac{\hat{p}}{2}\right) \;,
 \end{equation}
 where the variance $\text{Var}\left(\hat \rho^{B}_{\mathcal{I}_{k}|\phi},\hat{p}/2\right)$ in the conditional state $\hat \rho^{B}_{\mathcal{I}_{k}|\phi}$ \eqref{eq:IntervalCondState} is calculated in full analogy to \eqref{eq:var}.\\
 
\begin{figure}[!htb]
\centering
\includegraphics[width=0.46\textwidth]{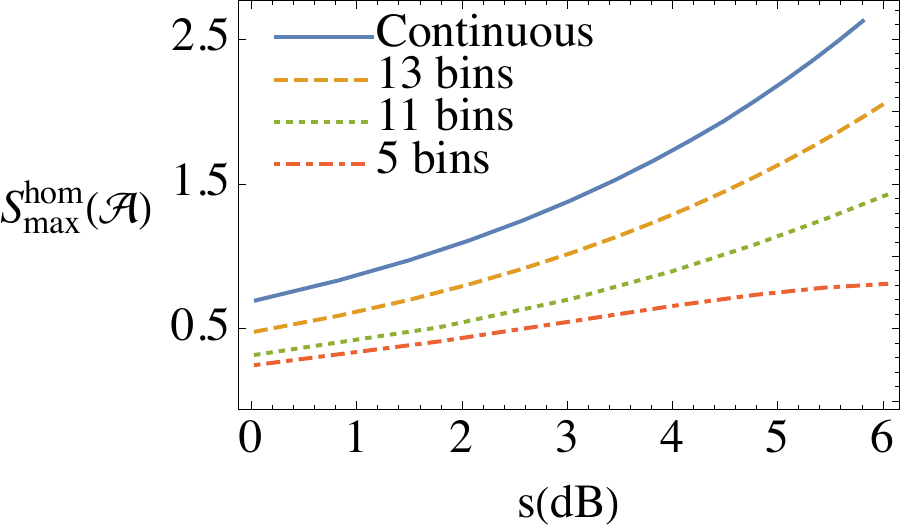}
\caption{Effect of discretization of the quadrature outcomes in Alice's side on the witness of steering for the two-mode photon-subtracted state with $\theta=\pi/4$ in the tuneable beam splitter in Fig.~\ref{fig:PhotonSubt_DiffBasis}. Below five bins it is not possible to witness steering as we fail to capture important features of the probability density. Nevertheless, a good violation of the inequality \eqref{eq:Metrological_Inequality} is possible starting from five bins.} 
\label{fig:witnessing_steering_Photon_subtracted_UnCorr_Basis_binned}
\end{figure}

 For the examples in Fig.~\ref{fig:witnessing_steering_Photon_subtracted_UnCorr_Basis_binned}, the (typically unequal) sizes of the different bins were optimized to maximize the witness. Because the photon-subtracted states have no mean field, we choose bins which are symmetric around the origin to reflect the structure of the exact quadrature statistics. We show the behavior of the steering witness \eqref{eq:Steering_witness} against the level of squeezing of the initial two-mode state, for the case of purely non-Gaussian steering corresponding to the choice $\theta=\pi/4$ in the second beamsplitter in Fig.~\ref{fig:PhotonSubt_DiffBasis}. Being able to witness steering in this challenging regime while considering realistic discretization of the measurement results, is particularly encouraging for the prospect of experimental implementations of this method.
 
\begin{figure}[!htb]
\centering
\includegraphics[width=0.46\textwidth]{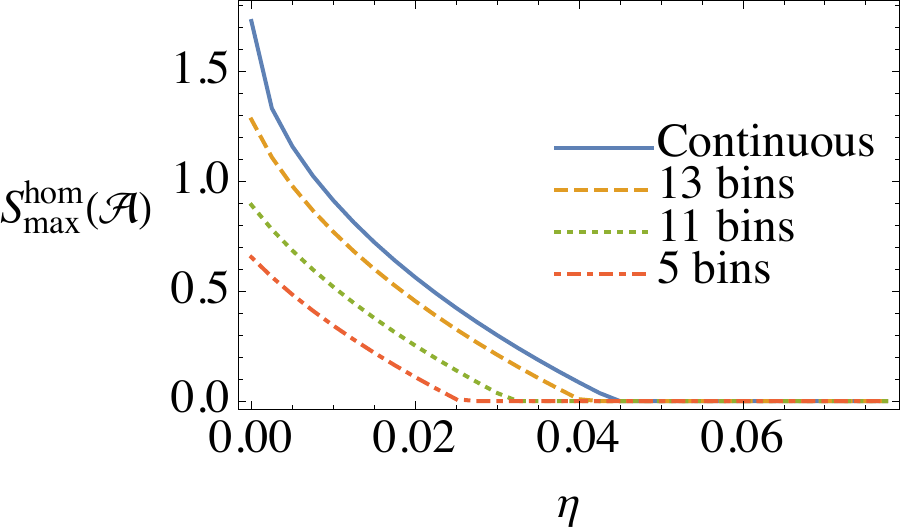}
\caption{Effect of losses in the witness of steering in the same kind of states as in Fig.~\ref{fig:witnessing_steering_Photon_subtracted_UnCorr_Basis_binned}, for a level of squeezing of $4$dB. As expected, the steering witness is reduced with increasing losses but again for a reasonably low number of bins, such as $13$, the resilience is almost the same as in the pure loss-free scenario.     } 
\label{fig:losses_steering_Photon_subtracted_UnCorr_Basis_bined}
\end{figure}

 In Fig.~\ref{fig:losses_steering_Photon_subtracted_UnCorr_Basis_bined} we analyze how binning the spectrum of outcomes of Alice's measurements affects the capability to witness the steering under the influence of photon losses. As expected, measurements with fewer bins, which lead to weaker violations of witness \eqref{eq:Metrological_Inequality}, also show a smaller tolerance to losses.
\subsection{Detecting quantum steering on homodyne data}
In this Section we present a \textit{realistic} analysis of the protocol that we have presented. So far, we have considered in an exact way the marginals of the Wigner functions. In an experimental implementation we would have to infer these probability densities from the outcomes of the homodyne measurements, or compute directly some of the quantities involved. To study such a scenario, we simulate experimental data through rejection sampling from the theoretical probability densities.

As we have observed so far, in the states that we have considered the largest violations are obtained when considering displacements along the $q$ or $p$ quadratures, conditioned on measurements in the same quadrature in Alice's side. This is particularly suited for an experimental implementation as simultaneous locking of the local oscillators in the phase and amplitude quadrature is already possible in homodyne detection schemes. Therefore, the data that we simulate for each measurement is sampled from the joint probability distribution of the same quadratures of both Bob's and Alice's modes, that can theoretically be obtained by integrating the Wigner function \eqref{eq:wigFinal} over the remaining quadratures. To better represent realistic experimental settings, the states that we consider will be slightly different from the ones that we analyzed before. In particular, the squeezing of the two modes will not be exactly the same. Thus, the choice of quadratures previously mentioned is not the optimal one, but it will always provide a lower bound for the actual value of the witness.

\begin{figure}[!htb]
\centering
\includegraphics[width=0.46\textwidth]{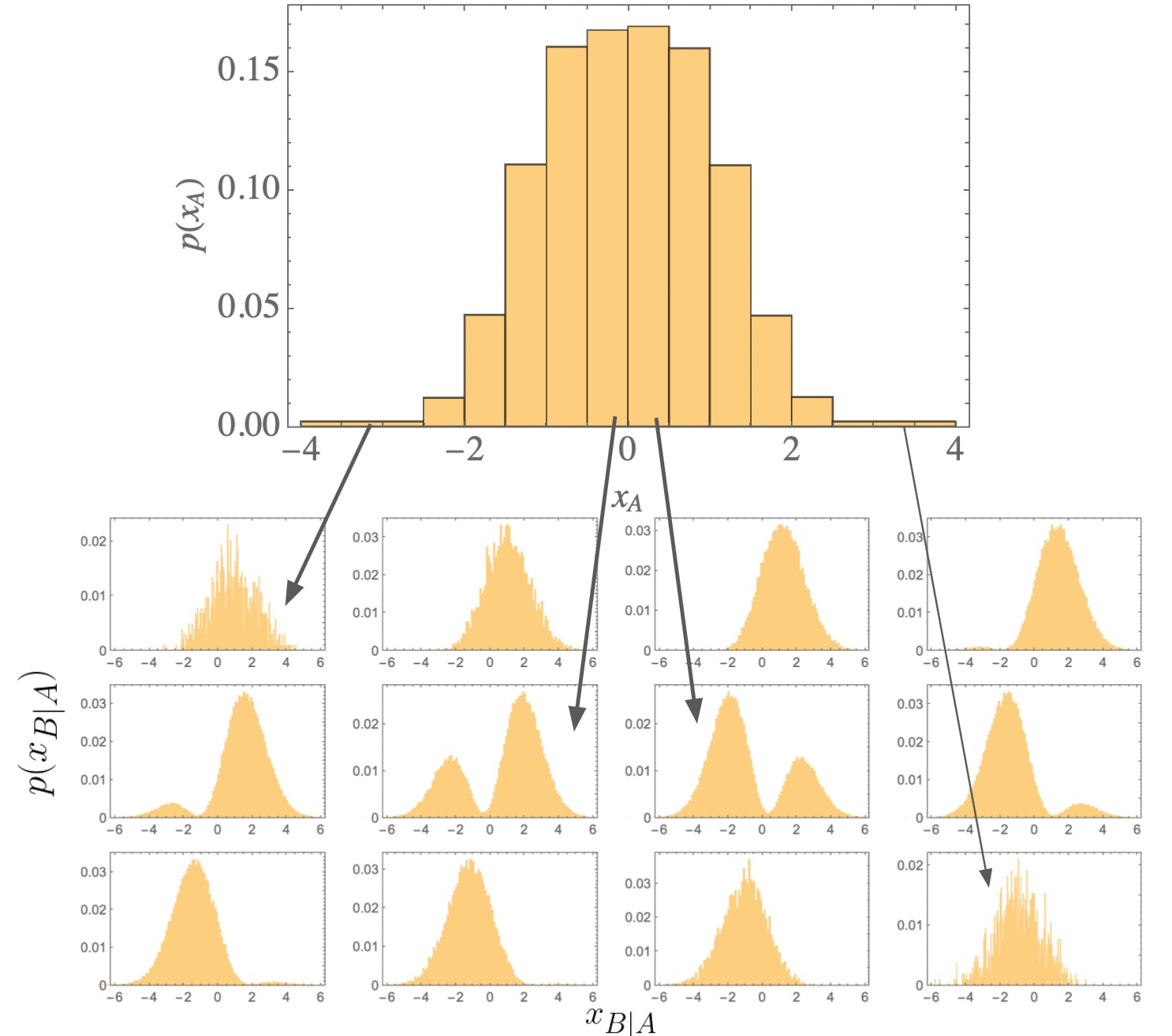}
\caption{Schematic representation of the analysis of the simulated data. The histogram in the top represents the outcome statistics of measurements of Alice's position quadrature. The panels below each represent the conditional statistics of the position quadrature of Bob's mode, corresponding to each of the bins of Alice's histogram. Wide tails, compared to the rest of the bins, are considered in order to guaranty sufficient statistics for the reconstruction of the states conditioned on less likely outcomes.} 
\label{fig:homodyne_data_analysis}
\end{figure}

In what follows, we discuss the protocol for the analysis of the data. Let us consider the simultaneous measurement of the momentum quadrature. The ideal reconstruction of the assemblage implies the inference of the probability density on Alice's side, and for each possible outcome, the reconstruction of the conditioned state. As mentioned in the previous subsection, this is an unfeasible experimental task, even more so if we consider the fact that one actually under-samples the tails of the distributions on Alice's side, in a way that reconstructing the statistics of its corresponding conditioned state is impossible. To overcome this issue we have to build a histogram on Alice's side, and analyze Bob's statistics conditioned on each bin of the histogram (Fig.~\ref{fig:homodyne_data_analysis}). It is important to remark that the histogram has to be inhomogeneous: the bins in the tails must encompass a larger region in order to avoid spurious contributions from under-sampled data. 

Computing the conditional variance is a rather straightforward task. On the other side, the computation of the FI from the discrete outcomes is more subtle. 
The most common procedure to experimentally estimate the FI relies on the computation of the Hellinger distance (statistical distance), between the reference probability density and the displaced ones \cite{Strobel2014,Wootters1981,Braunstein1994}.

\begin{figure}[t]
\centering
\includegraphics[width=0.25\textwidth]{Figures/Fig_A-B_Correlated.png}
\includegraphics[width=0.4\textwidth]{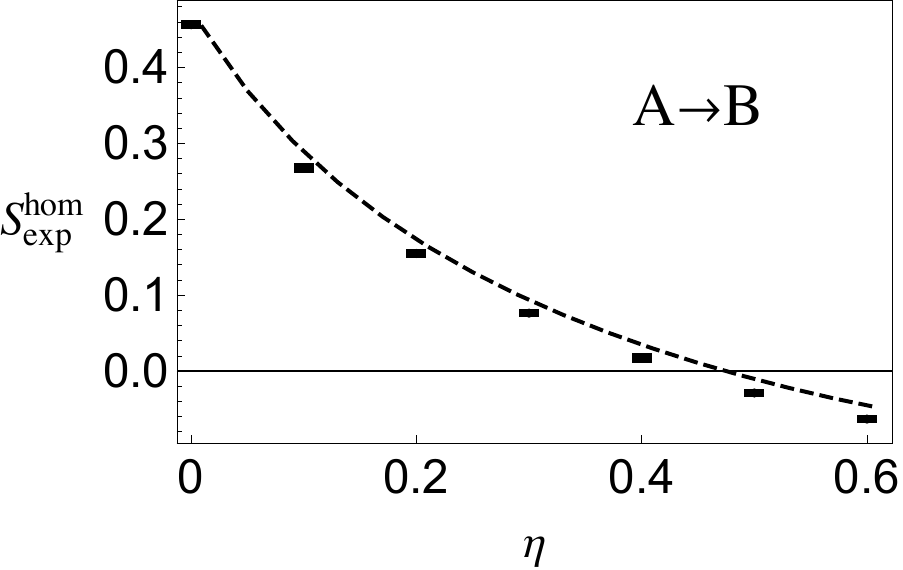}
\includegraphics[width=0.4\textwidth]{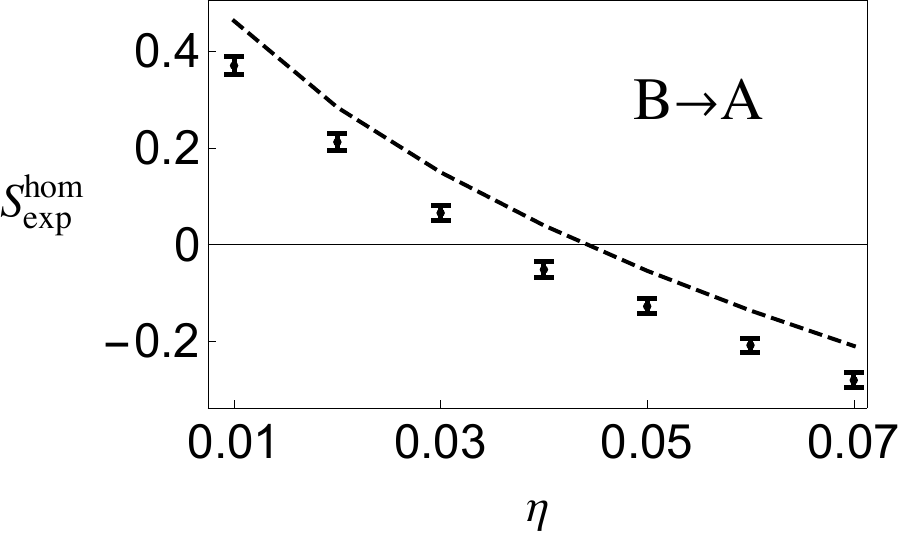}
\caption{Numerical simulations of the effect of losses in the witness of steering estimation  \eqref{eq:Steering_witness}. The state that we considered is defined by the choice of $\theta=0$ in the preparation scheme, \ie the state resulting after local photon subtraction in one of two correlated squeezed modes, with inhomogeneous squeezing $s_1=3.2 {\rm dB}$ and $s_2=2.6 {\rm dB}$. The plot in the top corresponds to the steering from the mode complementary to the one where the photon was subtracted. The bottom plot corresponds to the steering from the mode where the photon was subtracted. Dashed lines correspond to the exact analytical results obtained considering binning on the steering side. Error bars correspond to statistical errors and uncertainties on the fit. We can observe that the steering from the photon subtracted state, which is not observed using Gaussian criteria, vanishes somewhere between $3\%$ and $4\%$ losses, while the steering in the complementary direction persists for larger amount of losses. } 
\label{fig:dependence_losses-exp1}
\end{figure}

With a parameter-dependent probability density $P(q|\xi)$ and a reference $P(q|0)$, the Hellinger distance is defined as
\begin{equation}
    d_{H}^{2}(\xi)=\frac{1}{2}\int_{q} \left(\sqrt{P(q|\xi)}-\sqrt{P(q|0)}\right)^2dq.
\end{equation}
Expanding $P(q|\xi)$ to first order in $\xi$ it is possible to show that
\begin{equation}\label{eq:Hellinger_Dist_Fisher_Inf}
d_{H}^{2}(\xi)=\frac{F}{8} \xi^2 +\mathcal{O}(\xi^3),    
\end{equation}
where $F$ is shorthand for the Fisher information $F_{\xi = 0}[P]$.
Hence, it is enough to perform a quadratic fitting of the Hellinger distance to estimate the FI $F$. The latter is particularly well suited for our analysis as the displaced probability distributions can be obtained by just shifting the reference one. Such a post-processing displacement of the measurement outcomes does not require to experimentally implement the displacements, which is much more demanding. In an experimental implementation we have access to relative frequencies $\{\mathcal{F}(q|\xi)\}$ rather than the exact probabilities $P(q|\xi)$ required above.  In this context formula \eqref{eq:Hellinger_Dist_Fisher_Inf} is valid only as an approximation, given the fact that $\mathcal{F}(q|\xi)=P(q|\xi)+\delta \mathcal{F}(q|\xi)$, with $\delta \mathcal{F}(q|\xi)$ a statistical fluctuation that arises due to finite sample size.  Because of normalization $\sum_{q} \delta \mathcal{F}(q|\xi)=0$, where the sum runs over all possible values of $q$, which for CV systems will be given by all possible bins in which the outcome of the measurement might fall.  If we define the histograms $f(0)=\{\mathcal{F}(q|0)\}_q$ and $f(\xi)=\{\mathcal{F}(q|\xi)\}_q$ for a sample of $n$ experimental measurements, we have \cite{Strobel2014}
\begin{equation}
\langle d_{H}^{2}\left(f(0),f(\xi)\right)\rangle = c_0+\left(\frac{F}{8} + c_2\right) \xi^2 +\mathcal{O}(\xi^3,\delta \mathcal{F}(q|\xi)^3),
\end{equation} 
with
\begin{equation}
\begin{split}
&c_0=\frac{N-1}{4 n}\\
&c_2\approx \frac{F(1+N)}{32 n},
\end{split}
\end{equation}
where $\langle d_{H}^{2}\left(f(0),f(\xi)\right)\rangle$ is the sample average of the Hellinger distance between the two relative frequencies, $n$ is the number of measurements and $m$ is the number of values of $q$ for which $\mathcal{F}(q|\xi)\neq 0$. Observe that the previous formula converges asymptotically to \eqref{eq:Hellinger_Dist_Fisher_Inf}.  This implies that the estimation of $F$ is asymptotically unbiased, with the bias decreasing as $1/n$.\\

\begin{figure}
\centering
\includegraphics[width=0.45\textwidth]{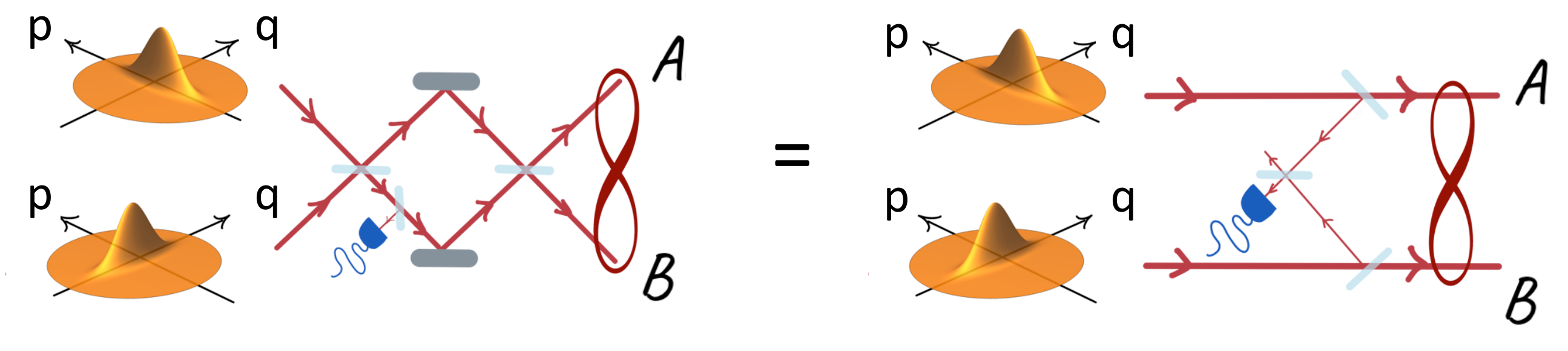}\\
\includegraphics[width=0.4\textwidth]{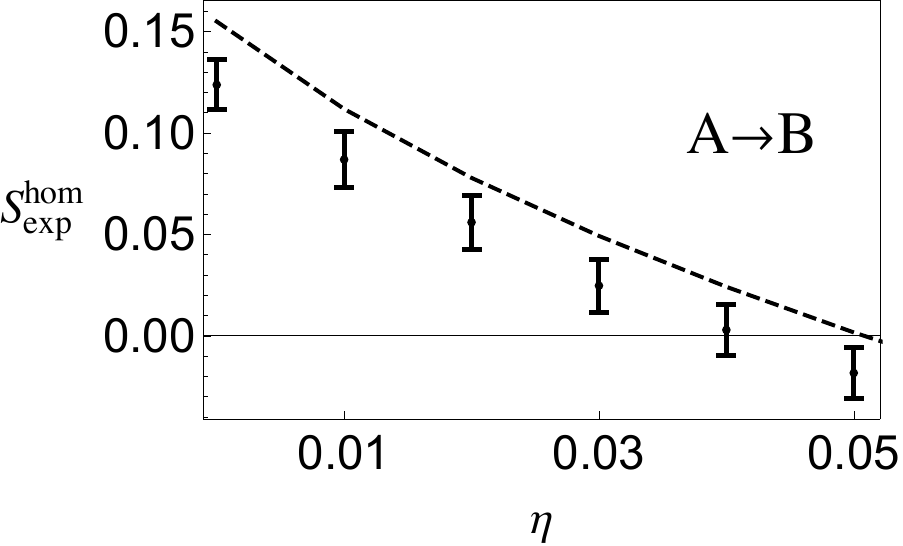}\\
\includegraphics[width=0.4\textwidth]{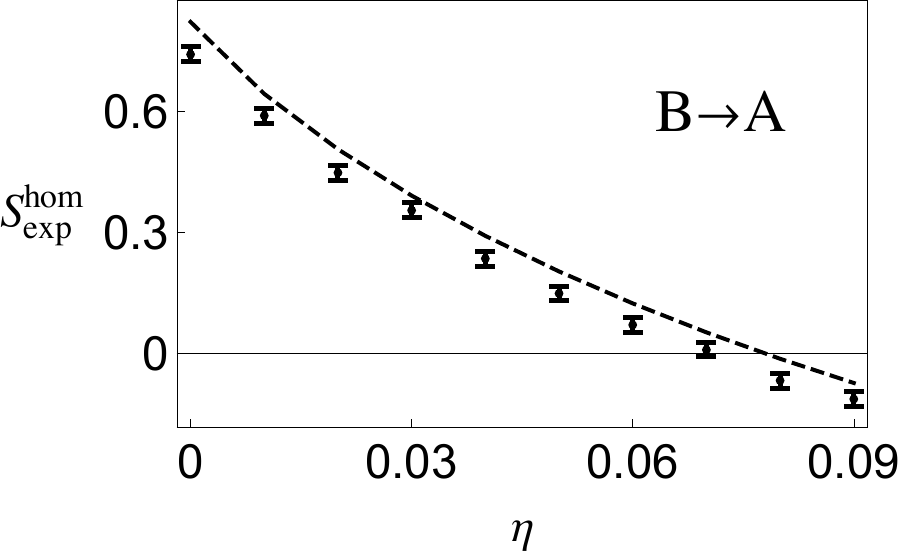}
\caption{Numerical simulations of the effect of losses in the witness of steering estimation \eqref{eq:Steering_witness}. The top panel represents the state preparation with $\theta=\pi/4$, \ie delocalized photon subtraction from a two-mode squeezed state with inhomogeneous squeezing $s_1=3.2$dB and $s_2=2.6$dB. Dashed lines correspond to the exact analytical results obtained considering binning on the steering side with a perfect measurement of the Fisher information (as in Subsection \ref{sec:Disc}), while points are obtained by inferring the Fisher information from the finite data using the Hellinger distance method (see text). Error bars correspond to statistical errors and uncertainties on the fit. Contrary to the results showed for homogeneous squeezing, here we can observe a remarkable difference in the steering in the two complementary directions.} 
\label{fig:dependence_losses-exp2}
\end{figure}

In Figs.~\ref{fig:dependence_losses-exp1} and \ref{fig:dependence_losses-exp2} we study the influence of losses on the estimation of the steering witness \eqref{eq:Steering_witness} based on a finite set of $n=10^5$ data points. In most experiments, it is unrealistic to have exactly the same squeezing in each mode. Therefore, the specific values of the squeezing in Alice's and Bob's initial states (at the left of Fig.~\ref{fig:PhotonSubt_DiffBasis}) are chosen arbitrarily. We chose  $s_1=3.2$db and $s_2=2.6$db since these lie in an experimentally relevant range. 
The obtained violation is below the exact result $S_{\text{max}}^{\text{hom}}$ (dashed line) obtained when considering the exact Wigner function of the system for the same set of parameters. 

In Fig.~\ref{fig:dependence_losses-exp1}, we investigate this system in the correlated basis, i.e., the cases $\theta = 0$, where Bob subtracts a photon. Subsequently, we analyse the steering from Alice to Bob and from Bob to Alice. As before, the case where Alice steers Bob can be studied using Reid's criterion, as shown in Fig. \ref{fig:witnessing_steering_Photon_subtracted_Corr_Basis}(a), indicating that it is essentially a case of Gaussian steering. Also in these simulations, based on a finite number of data points, this Gaussian character translates to a much greater resilience against losses. However, for steering from Bob to Alice --which cannot be witnessed through Reid's criterion-- we observe a much more detrimental effect of losses. Because non-Gaussian features are typically very sensitive to losses, a possible origin of this sensitivity is that the steering is dominated by non-Gaussian features of the state. This conjecture is supported by Fig.~\ref{fig:losses_steering_Photon_subtracted_Corr_Basis}, which shows that the protocol is much more loss-resistant when the correlations have some Gaussian features.

To fully explore the feasibility of our method for witnessing non-Gaussian steering, we show the case for $\theta = \pi/4$ in Fig.~\ref{fig:dependence_losses-exp2}. In this scenario, all the correlations in the state (be it quantum or classical) originate from the non-Gaussian part of the Wigner function \eqref{eq:wigFinal} and no quantum correlation can be witnessed based on its covariance matrix. In other words, this is a state where all quantum steering is purely non-Gaussian in nature. Again, we observe a much more detrimental effect of losses compared to the top panel of Fig.~\ref{fig:dependence_losses-exp1}. However, due to the asymmetry in the steering of the two modes, we observe a much larger value for the steering witness when considering steering from the lesser squeezed mode to the more squeezed modes. This higher value also comes with a higher robustness to losses. From an experimental point of view, the tolerable loss values remain very small in both cases. Nevertheless, these simulations show that in sufficiently pure systems it is possible to witness non-Gaussian quantum steering using exclusively homodyne detection with an experimentally feasible protocol.


\section{\label{sec:Conclusions} Conclusions and outlook}
We proposed a protocol for witnessing steering in CV systems. The protocol is based on the metrological steering criterion first proposed in Ref.~\cite{EPR_Metr2021}, and relies solely on homodyne detection. The latter makes it suitable for current experimental capabilities. The protocol is shown to succeed in detecting quantum steering in non-Gaussian states, even in scenarios where protocols based on Gaussian features, like Reid's criterion, are shown to fail, when restricted to quadrature measurements. A comparison between our metrological protocol and the entropic witness presented in \cite{PhysRevLett.106.130402} shows that our protocol consistently outperforms the entropic one. This adds to a similar conclusion that was reached in \cite{PhysRevA.94.020101} for a comparison between metrological and entropic entanglement witnesses. It remains an interesting open question whether there is a formal way of proving that the metrological witness is always larger than the entropic one. Such a proof could potentially lead to new insights in the relation between the Fisher information and entropy.

A realistic simulation of data from a continuous-variable experiment includes the effects of loss, data discretization and the scalable extraction of the Fisher information. Our results show that non-Gaussian quantum steering can be detected with a feasible number of measurements. Even for reasonably small numbers of samples ($n=10^5$), the violation of the inequality \eqref{eq:Metrological_Inequality} can be observed with several standard deviations, considering around 3dB of squeezing, albeit requiring rather low losses. Rather than a feature of our specific protocol, the high sensitivity to losses for these states might be an indication of the fragile nature non-Gaussian quantum steering. We should emphasise that our metrological witness is based on the same experimental implementation as Reid's criterion for quadrature operators. However, the post-processing of the measurement data is significantly more involved in our approach.

The relevance of our protocol is not merely experimental. Non-Gaussian quantum correlations are notoriously difficult to study in CV systems. For the most complete descriptions of CV states, one generally resorts to quasi-probability distributions. However, it is highly challenging to use such objects to study quantum correlations (Bell inequalities are a notable exception \cite{PhysRevLett.82.2009,PhysRevA.74.052114}). The techniques in Section \ref{sec:ProtocolCV} provide a useful way to analytically study the presence of metrologically useful non-Gaussian quantum steering based purely on the marginal of the Wigner function. 


\begin{acknowledgements}
This work was supported by the ANR JCJC project NoRdiC (ANR-21-CE47-0005).
It was also partially funded by the European Union’s Horizon 2020 research and innovation programme under Grant Agreement No. 899587, the Marie Sk\l{}odowska-Curie Grant Agreement No. 847648 with fellowship code LCF/BQ/PI21/11830025, and the QuantERA programme through the project ApresSF. M.G. acknowledges the LabEx ENS-ICFP: ANR-10-LABX-0010/ANR-10-IDEX-0001-02 PSL* and the Ministerio de Ciencia e Innovaci\'{o}n (MCIN) / Agencia Estatal de Investigaci\'on (AEI) for Project No. PID2020-115761RJ-I00 and support of a fellowship from ``la Caixa” Foundation (ID 100010434). This work was supported within the QuantERA II Programme that has received funding from the European Union’s Hori- zon 2020 research and innovation programme under Grant Agreement No 101017733.
\end{acknowledgements}

\bibliography{References}

\onecolumngrid 
\appendix

\section{Genuinely multimode protocol}\label{app:MultimodeMode}

The protocol that was proposed in Section \ref{sec:ProtocolCV} effectively describes a witness for steering between two modes in a larger multimode system. In this section, we provide an extension of the protocol in a more general multimode setting. The resulting steering witness is strictly better for testing the steering between Alice and Bob, but it comes with considerably more experimental overhead and parameters to optimise.\\

First of all, let us consider Alice's subsystem which contains $m$ modes. Rather than just choosing one axis $\vec f$ in Alice's phase space along which to measure, we can choose any set of axes that are not connected to the same mode. Formulated differently, for any mode basis in Alice's subsystem, we can measure one quadrature in each mode and condition on the joint outcome for all these measurements. 

To formalise this idea, let us first consider an orthonormal symplectic basis ${\cal F}$ of Alice's phase space, given by
\begin{equation}\label{eq:AliceBasisMulti}
    {\cal F} = \{\vec f_1, \Omega f_1, \dots, \vec f_m, \Omega \vec f_m \},
\end{equation}
where $\vec f_k$ are all vectors in $\mathbb{R}^{2M},$ where $M$ is the number of modes in the global system that contains both Alice and Bob. One can think of ${\cal F}$ as one of the infinitely many ways of identifying axes in Alice's phase in a way such that $\vec f_k$ and $\Omega f_k$ always belong to the same mode (we could say that the axis generated by $\vec f_k$ represents the measurements of the $q$-quadrature in this mode and $\Omega f_k$ generates the axis that represents its $p$-quadrature).

The vectors $\vec f_1, \dots, \vec f_m$ now by construction correspond to $m$ axes in phase space that can be jointly measured. When we perform such a measurement and post-select on a series of measurement outcomes $x_1, \dots, x_m$ for each one of these axes, we find that Bob's Wigner function is transformed into
\begin{equation}\label{eq:WignerCondMultimode}
\begin{split}
    &W^{B|A}(\vec{x}_{B}|x_{A}^{\vec f_1}=x_1, \dots, x_{A}^{\vec f_m}=x_m)=\frac{\int_{\Re^{2m}} W(\vec{x}_{A}\oplus \vec{x}_{B}) \prod_{k=1}^m\delta(\vec f_k^{\top}\vec x_A - x_k) d\vec{x}_{A}}{P_A(x_{A}^{\vec f_1}=x_1, \dots, x_{A}^{\vec f_m}=x_m)},
    \end{split}
\end{equation}
where we define
\begin{equation}\begin{split}
P_A&(x_{A}^{\vec f_1}=x_1, \dots, x_{A}^{\vec f_m}=x_m)=\int_{\Re^{2m}\oplus \Re^{2m'}}W(\vec{x}_{A}\oplus \vec{x}_{B}) \prod_{k=1}^m\delta(\vec f_k^{\top}\vec x_A-x_k) d\vec{x}_{A} d\vec{x}_{B}.
\end{split}
\end{equation}
Eq. \eqref{eq:WignerCondMultimode} thus directly generalises \eqref{eq:WignerCond}.\\

On Bob's side of the system, we are now going to use this Wigner function to study the effect of a change in mean field. One particular feature of such displacement operations is that they are generated by a quadrature operator, which means that they are always acting along a well-defined axis $\vec e$ in Bob's phase space. The parameter of interest thus affects Bob's conditional state as
\begin{equation}\begin{split}
    &W^{B|A}(\vec{x}_{B}|x_{A}^{\vec f_1}=x_1, \dots, x_{A}^{\vec f_m}=x_m)\mapsto W^{B|A}(\vec{x}_{B} - \xi \vec e|x_{A}^{\vec f_1}=x_1, \dots, x_{A}^{\vec f_m}=x_m).
    \end{split}
\end{equation}
Because on Bob's side we implement a parameter with a single-mode generator, the calculation of the conditional variance generalises in a straightforward fashion:
\begin{equation}\label{eq:VarCondMulti}
\text{Var}_{\rm hom}^{B|A}\left(\mathcal{A},\frac{ \vec e^{\top} \Omega \vec{\hat x}}{2}\right)=\min_{{\cal F}}\frac{1}{4}\int_{\Re^m}P_A(x_{A}^{\vec f_1}=x_1, \dots, x_{A}^{\vec f_m}=x_m) \text{Var}\left(\hat \rho_{x_1, \dots x_m \lvert \vec f_1, \dots \vec f_m}^{B},\vec e^{\top} \Omega \vec{\hat x}\right)dx_1\dots dx_m,
\end{equation}
where $\text{Var}\left(\hat \rho_{x_1, \dots x_m \lvert \vec f_1, \dots \vec f_m}^{B},\vec e^{\top} \Omega \vec{\hat x}\right)$ is the variance of the quadrature corresponding to the generator $\vec e^{\top} \Omega \vec{\hat x}$. To compute this quantity, we use exactly the same subspace of phase space as before, generated by all vectors orthogonal to $\Omega \vec e$:
\begin{equation}
 {\cal P}^{\perp} = \{\vec x_B \in \Re^{2m'}  \mid \vec e^{\top} \Omega \vec x_B = 0 \}.   
\end{equation}
We then calculate the measurement statistics for the quadrature $\vec e^{\top} \Omega \vec{\hat x}$ as
\begin{equation}\label{eq:cond_mom_prob_densMulti}
    \tilde{P}^{B}_{x_1, \dots, x_m \lvert \vec f_1, \dots \vec f_m}(p)=\int_{ {\cal P}^{\perp}}  W^{B|A}(\vec{x}_{B}|x_{A}^{\vec f_1}=x_1, \dots, x_{A}^{\vec f_m}=x_m) d\vec{x}^{B},
\end{equation}
where $p$ denoted the values along the single remaining phase space axis generated by $\vec e^{\top} \Omega \vec{\hat x}$.
This distribution allows us to compute
\begin{equation}\label{eq:varMulti}
\text{Var}\left(\hat \rho_{x_1,\dots, x_m \lvert \vec f_1, \dots, \vec f_m }^{B},\vec e^{\top} \Omega \vec{\hat x}\right) =\int_{\Re} p^2 \tilde{P}^{B}_{x_1, \dots, x_m \lvert \vec f_1, \dots \vec f_m}(p)\, dp - \left(\int_{\Re} p \tilde{P}^{B}_{x_1, \dots, x_m \lvert \vec f_1, \dots \vec f_m}(p)\, dp\right)^2
\end{equation}
In practice, this is still the variance of only one quadrature operator in Bob's conditional state. From an experimental point of view, this can be considered a significant advantage due to limited overhead.\\

The biggest difference appears on the level of the Fisher information. In Eq. \eqref{eq:class_FI_hom}, we only use the specific displaced quadrature along the phase space axis $\vec e$. However, more generally speaking, we can use any set of quadratures in Bob's subsystem to estimate the displacement strength $\xi$. To formalise this idea, we are going to consider the case where we use $m'$ (the number of modes in Bob's subsystem) jointly measurable quadratures to estimate $\xi$. To do so, we will use the Wigner function \eqref{eq:WignerCondMultimode} and integrate out all the complementary quadratures. To maximize the efficiency of the parameter estimation, we will always consider cases where the full displacement is contained within the set of quadratures that is used to estimate it. 

For this purpose, let us introduce a symplectic orthonormal basis ${\cal G}$ of Bob's phase space $\mathbb{R}^{2m'}$ :
\begin{equation}\label{eq:BobBasisPhaseSpaceMulti}
    {\cal G} = \{\vec g_1, \Omega \vec g_1, \dots, \vec g_{m'}, \Omega \vec g_{m'} \}.
\end{equation}
A crucial additional constraint that is imposed on this basis is that some $\alpha_k \in \mathbb{R}$ with $\sum_k \alpha_k^2 = 1$ exist such that
\begin{equation}
    \vec e = \alpha_1 \vec g_1 + \dots + \alpha_{m'} \vec g_{m'}.
\end{equation} 
This demand is important, because we are going to measure quadratures along the phase space axes generated by $\vec g_1, \dots, \vec g_{m'}$. When doing so, we generalise the expression \eqref{eq:wahwah} to
\begin{equation}\label{eq:wahwahMulti}
    P^{B}_{x_1, \dots, x_{m} \lvert \vec f_1, \dots, \vec f_m}(q_1, \dots, q_{m'})=\int_{ \mathbb{R}^{2m'}} \prod_{k=1}^{m'} \delta( \vec g_k^{\top}\vec{x}_B - q_k)W^{B|A}(\vec{x}_{B}|x_{A}^{\vec f_1}=x_1, \dots, x_{A}^{\vec f_m}=x_m) d\vec{x}_{B}.
\end{equation}
The action of the displacement now becomes a bit more subtle, in the sense that
\begin{equation}
P^{B}_{x_1, \dots, x_{m} \lvert \vec f_1, \dots, \vec f_m}(q_1, \dots, q_{m'} \mid \xi) = P^{B}_{x_1, \dots, x_{m} \lvert \vec f_1, \dots, \vec f_m}(q_1 - \alpha_1 \xi, \dots, q_{m'} - \alpha_{m'}\xi \mid \xi)
\end{equation}
The Fisher information for estimating $\xi$ using this multivariate distribution can be calculated by a straightforward extension of \eqref{eq:Classical_FI}, such that we find
\begin{equation}\label{eq:Classical_FI_multi}
    F^B_{\xi}[P^{B}_{x_1, \dots, x_{m} \lvert \vec f_1, \dots, \vec f_m}] = \int_{\Re^{m'}} P^{B}_{x_1, \dots, x_{m} \lvert \vec f_1, \dots, \vec f_m}(q_1, \dots, q_{m'} \mid \xi) \left(\frac{\partial \mathcal{L}(q_1, \dots, q_{m'} \mid \xi) }{\partial \xi}\right)^2 d q_1\dots dq_{m'}.
\end{equation}
The conditional Fisher information then becomes \begin{equation}\label{eq:class_FI_hom_Multi}
    F^{B|A}_{\rm hom}\left(\mathcal{A},\frac{\vec e^{\top} \Omega \vec{\hat x}}{2}\right)=\max_{{\cal F} }\int_{\Re^m} P_A(x_A^{\vec f_1}=x_1, \dots, x_A^{\vec f_m}=x_m) F^B_{\xi}[P^{B}_{x_1, \dots, x_{m} \lvert \vec f_1, \dots, \vec f_m}]dx_1 \dots dx_m.
\end{equation}
Note that we maximize over all possible bases for Alice's phase space ${\cal F}$, as given by \eqref{eq:AliceBasisMulti}.\\

Combining all the above elements now leads us to formulate a fully multimode version of the metrological witness \eqref{eq:Steering_witness_hom}:
\begin{equation}\label{eq:Steering_witness_hom_Multi}\begin{split}
    S_{\text{max}}^{\text{hom}}(\mathcal{A})=\max_{\vec e \in \Re^{2m'}; {\cal G}}\left[F^{B|A}_{\rm hom}\left(\mathcal{A},\frac{\vec e^{\top} \Omega \vec{\hat x}}{2}\right)-\text{Var}_{\rm hom}^{B|A}\left(\mathcal{A},\vec e^{\top} \Omega \vec{\hat x}\right)\right]^{+},
    \end{split}
\end{equation}
where the terms are now defined through \eqref{eq:VarCondMulti} and \eqref{eq:class_FI_hom_Multi}. Furthermore, we note that we must maximise this value over all possible choices of displacement directions and subsequently all the possible ways of constructing a basis ${\cal G}$ of Bob's phase space according to \eqref{eq:BobBasisPhaseSpaceMulti}. Of course, in practice any displacement direction and measurement basis that allows to obtain a value of $S_{\text{max}}^{\text{hom}}(\mathcal{A})$, that is significantly larger than zero (significant as compared to an experimental error bar), is sufficient to certify quantum steering form Alice to Bob.\\

The steering witness in \eqref{eq:Steering_witness_hom_Multi} is guaranteed to outperform the version in \eqref{eq:Steering_witness_hom} in which Bob only measures the displaced quadrature. However, it is clear that having to optimise several homodyne detectors to function simultaneously clearly requires much more experimental overhead than using a single detector. This thus imposes the question whether there is a strict advantage in using  the multimode witness \eqref{eq:Steering_witness_hom_Multi}, where Alice and Bob measure all their quadratures simulataneously.

For Alice's measurements we explore the case where no individual mode (regardless of the mode basis) can steer Bob, but where we require the use of several modes at the same time. On Bob's side, the matter is more related to metrology. Because the displacement is anyway generated by a generator that acts on one specific mode, it is logical to wonder whether only measuring the displaced quadrature operator is sufficient to extract all information on $\xi$. There is an argument to suggest that this is typically not the case. When in the state given by \eqref{eq:WignerCondMultimode} the mode in which the displacement acts is entangled to other modes, a measurement of only the displaced quadrature will trace out the other modes which effectively leads to decoherence. This suggests that in these cases \eqref{eq:Steering_witness_hom_Multi} could detect steering that remains hidden when the simpler form \eqref{eq:Steering_witness_hom} is used. This can be verified by comparing the obtain FI to the QFI if the latter can be calculated.

A detailed study of all these extra effects would require us to perform additional case studies for different kind of multimode states. However, such a study requires a more dedicated effort and is considered to be beyond the scope of this work.

\end{document}